\documentclass[prx,twocolumn,showpacs,showkeys,superscriptaddress,preprintnumbers,floatfix,amsmath,amssymb,longbibliography]{revtex4-1}

\usepackage{graphicx}
\usepackage{dcolumn}
\usepackage{bm}
\usepackage{multirow}
\usepackage{array}

\usepackage[usenames,dvipsnames]{xcolor}
\definecolor{nblue}{rgb}{0.0, 0.0, 1.0}
\definecolor{magenta}{rgb}{0.79, 0.08, 0.48}
\usepackage[colorlinks,linkcolor=nblue,urlcolor=magenta,citecolor=magenta,plainpages=false,pdfpagelabels,breaklinks]{hyperref}
\usepackage{changes}

\newcommand{\beq}{\begin{equation}}
\newcommand{\eeq}{\end{equation}}
\newcommand{\bea}{\begin{eqnarray}}
\newcommand{\eea}{\end{eqnarray}}

\DeclareUnicodeCharacter{2062}{}
\begin{document}

\title{Two-stage evolution of magnetic correlations in spiral spin liquid material, Ca$_{10}$Cr$_{7}$O$_{28}$}

\author{Changhyun Koo}
\address{Kirchhoff Institute of Physics, Heidelberg University, Heidelberg 69120, Germany}
\address{Department of Physics, Sungkyunkwan University, Suwon 16419, Republic of Korea}

\author{Jaena Park}
\address{Kirchhoff Institute of Physics, Heidelberg University, Heidelberg 69120, Germany}

\author{Johannes Werner}
\address{Kirchhoff Institute of Physics, Heidelberg University, Heidelberg 69120, Germany}

\author{Suheon Lee}
\affiliation{Center for Artificial Low Dimensional Electronic Systems, Institute for Basic Science, Pohang 37673, Republic of Korea}

\author{Christian Balz}
\address{Helmholtz-Zentrum Berlin f\"{u}r Materialien und Energie, Berlin 14109, Germany}
\address{Institut f\"{u}r Festk\"{o}rperphysik, Technische Universit\"{a}t Berlin, Berlin 10623, Germany}

\author{A.T.M. Nazmul Islam}
\address{Helmholtz-Zentrum Berlin f\"{u}r Materialien und Energie, Berlin 14109, Germany}

\author{Yugo Oshima}
\address{RIKEN Cluster for Pioneering Research, Wako, Saitama 351-0198, Japan}

\author{Bella Lake}
\address{Helmholtz-Zentrum Berlin f\"{u}r Materialien und Energie, Berlin 14109, Germany}
\address{Institut f\"{u}r Festk\"{o}rperphysik, Technische Universit\"{a}t Berlin, Berlin 10623, Germany}

\author{Kwang-Yong Choi}
\email[]{choisky99@skku.edu}
\affiliation{Department of Physics, Sungkyunkwan University, Suwon 16419, Republic of Korea}

\author{R\"{u}diger Klingeler}
\email[]{klingeler@kip.uni-heidelberg.de}
\address{Kirchhoff Institute of Physics, Heidelberg University, Heidelberg 69120, Germany}

\newcommand{\ck}[1]{\textcolor{red}{#1}}

\date{\today}

\begin{abstract}
We present an X-band and tunable high-frequency/high-field electron spin resonance (HF-ESR) study of single-crystalline Ca$_{10}$Cr$_{7}$O$_{28}$, which constitutes alternating antiferromagnetic and ferromagnetic kagome bilayers. At high temperatures, a phonon-assisted relaxation process is evoked to account for the pronounced increase of the linewidth in an exchange-narrowing regime ($k_{\rm B}T\gg J$). In contrast, at low temperatures ($k_{\rm B}T\lesssim J$), a power-law behavior in line narrowing is observed. Our data reveal two distinct power-law regimes for the linewidth which crossover at $T^*\approx 7.5$~K. Notably, the intriguing evolution of the ESR linewidth in this alternating kagome bilayer system with opposite sign of exchange interactions highlights distinct spin dynamics compared to those in a uniform kagome antiferromagnet. 
\end{abstract}

\maketitle

\section{Introduction}

The search for quantum spin liquids (QSLs) has been a central theme in contemporary condensed matter physics. In QSLs, many-body interacting spins are long-range entangled and form a coherently fluctuating, liquid-like ground state without any spontaneously broken symmetry. The defining features of QSLs are no static magnetism, emergent topological order, and fractionalized quasiparticles~\cite{Balents,Wen}. While in the $S=1/2$ Heisenberg antiferromagnetic chain, which may be considered as a one-dimensional (1D) analog of a QSL, spinon excitations carrying a fractional quantum number have been well established~\cite{Lake,Mourigal}, experimental evidence is less clear in higher-dimensional QSL-candidate materials~\cite{BroholmScience2020}. Key ingredients to stabilize both gapless and gapped QSLs can be geometrical and exchange frustration. Frustration provides a prime route to generate macroscopic degeneracy of ground states, thereby, suppressing static long-range order through strong quantum fluctuations among possible competing spin configurations. Prototypical examples of geometrically frustrated magnets are triangular, kagome, and pyrochlore lattices~\cite{Shimizu,Shen,Pratt,Lee,Yan,Han,Gingras,Li,Xu,Xu23}, whilst exchange frustration is nicely embodied in the Kitaev honeycomb lattice~\cite{Kitaev,Baskaran,Knolle}.

Going forward, it is highly desirable to search for new classes of QSL materials which realize a frustration motif. In this regard, the recently discovered QSL compound Ca$_{10}$Cr$_{7}$O$_{28}$ could advance our understanding of QSLs as its Hamiltonian is distinct from the canonical spin liquid models mentioned above~\cite{Balz1,Balz2,Balz3,Balodhi,Pohle,Ni,Sonnenschein,Kshetrimayum}. The crystal structure of Ca$_{10}$Cr$_7$O$_{28}$ comprises breathing kagome bilayers of Cr$^{5+}$ ions ($S=1/2$) lying in the $ab$-plane and stacked along the $c$-axis in the space group of trigonal $R3c$~\cite{Balz3}. As sketched in Fig. 1, each kagome layer consists of two alternating corner-sharing equilateral triangles with distinct exchange interactions: one triangle exhibits ferromagnetic (FM) coupling while the spins of the other triangle are antiferromagnetically (AF) coupled~\cite{Balz1}. The magnetic interactions in these two layers are different. FM triangles with $J_{22} = -0.27(3)$~meV are stacked on top of AF ones with $J_{32} = 0.11(3)$~meV, while AFM triangles with $J_{31} = 0.09(2)$~meV are stacked on top of FM ones with $J_{21}= -0.76(5)$~meV. The stacked triangles are coupled to each other by a weak ferromagnetic intra-bilayer coupling, $J_{0} = -0.08(4)$~meV. The bilayers are decoupled from each other and the system is effectively two-dimensional~\cite{Balz1,note}. Noticeably, the vertex-to-vertex arrangement of the alternating FM and AF triangles both in the $ab$-plane and between the two layers of the bilayers results in a macroscopic degeneracy, potentially stabilizing a QSL ground state~\cite{Pohle}.

\begin{figure}
\label{figure1}
\includegraphics[width=8.5cm]{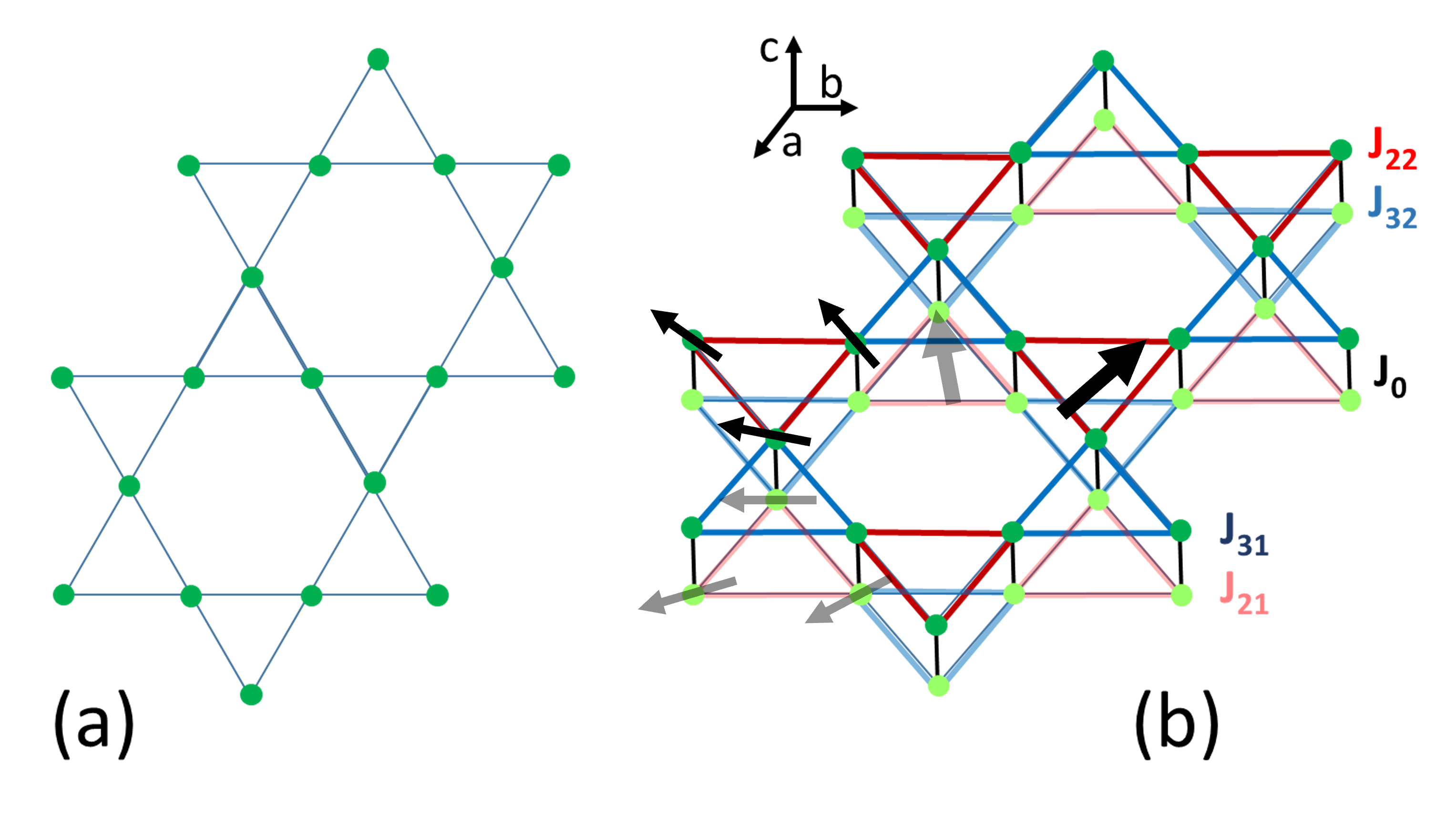}
\caption{Schematic of conventional uniform kagome structure (a) and alternating kagome bilayer structure (b). Blue and red triangles represent FM and AFM couplings within the triangles, respectively. Small arrows represent effective spins in a $S=1/2$ kagome-lattice structure and big arrows represent effective spins in a $S=3/2$ Heisenberg honeycomb lattice structure.}
\end{figure}

Indeed, magnetic heat capacity, ac susceptibility, and $\mu$SR studies on Ca$_{10}$Cr$_7$O$_{28}$ show neither long-range magnetic order nor spin freezing down to 19 mK~\cite{Balz1}. Rather, a dynamically fluctuating spin state is evidenced by the observation of persisting spin dynamics down to the lowest temperatures by $\mu$SR. In addition, inelastic neutron scattering (INS) shows two broad diffuse bands in the energy ranges of $0-0.6$~meV and $0.7-1.6$~meV, with hexagonal ring-like scattering, indicative of spinon excitations~\cite{Balz1,Balz2,Sonnenschein}. The spinon character of these excitations is further supported by low-temperature specific heat and heat transport data which show that the spectrum is either gapless or exhibits only a very small gap~\cite{Sonnenschein,Ni}. All these features are consistent with the formation of a QSL ground state in Ca$_{10}$Cr$_7$O$_{28}$. Additionally, recent spin noise measurements, combined with Monte Carlo simulations, suggest that Ca$_{10}$Cr$_7$O$_{28}$ hosts a spiral spin liquid state~\cite{Blundell}. This raises the question of whether spin correlations in the FM/AF alternating kagome bilayer system at hand differ from those in uniform AF kagome single layers as suggested by a recent numerical study of Pohle \textit{et al}~\cite{Pohle}. Their semi-classical simulations indicate that the QSL ground state incorporates two different types of spin correlations at two distinct time-scales: one associated with fluctuations inherent to a kagome–lattice antiferromagnet of localized spins $S=1/2$, and the other described by an effective $S=3/2$ Heisenberg honeycomb lattice~\cite{Pohle} (Fig. 1). 

Electron spin resonance (ESR) provides fundamental information about spin dynamics and magnetic anisotropies and is an experimental method of choice to investigate the evolution of spin correlations in the precursor phase of a QSL ground state. In this work, we hence present an X-band and tunable high-frequency (HF) ESR study of Ca$_{10}$Cr$_7$O$_{28}$ above the critical temperature of the spiral QSL phase. Our ESR spectra show no indication of critical slowing down of spin dynamics or gapped low-energy excitations. The salient finding is temperature dependence of the ESR linewidth, which follows two distinct power-law behaviors below 20~K. This result implies the existence of two different spin-correlation regimes inherent to the alternating bilayer kagome lattice of Ca$_{10}$Cr$_7$O$_{28}$.

\section{Experiments}

Single crystals (2 $\times$ 3 $\times$ 0.2 mm$^{3}$) of Ca$_{10}$Cr$_7$O$_{28}$ were grown as described in Ref.~\onlinecite{Balz3}. The samples employed for our ESR measurements were characterized by means of dc susceptibility, magnetization, heat capacity, INS, and heat transport in the earlier studies~\cite{Balz1,Balz2,Balz3,Balodhi,Ni,Sonnenschein}. X-band ESR spectra were obtained by means of a JEOL spectrometer at a fixed frequency of $\nu=9.12$~GHz and in magnetic fields of $\mu_0 H=0 - 1$~T. The spectra were recorded in the temperature range $T =3.6 - 300$~K. HF-ESR  measurements were performed using a transmission-type probe in a wide frequency range of $50 - 190$~GHz, and at temperatures between $T=2$~K and 140~K. Temperature control was ensured by temperature sensors in both the probe and sample space which was placed in the Variable Temperature Insert (VTI) of an Oxford magnet system equipped with a $16$~T superconducting coil. A phase-sensitive millimeter-wave vector network analyzer (MVNA) from AB Millim\`{e}tre was used as a source and detector~\cite{Comba2015}.

\section{Results and Discussion}

\begin{figure}
\label{figure2}
\includegraphics[width=8.5cm]{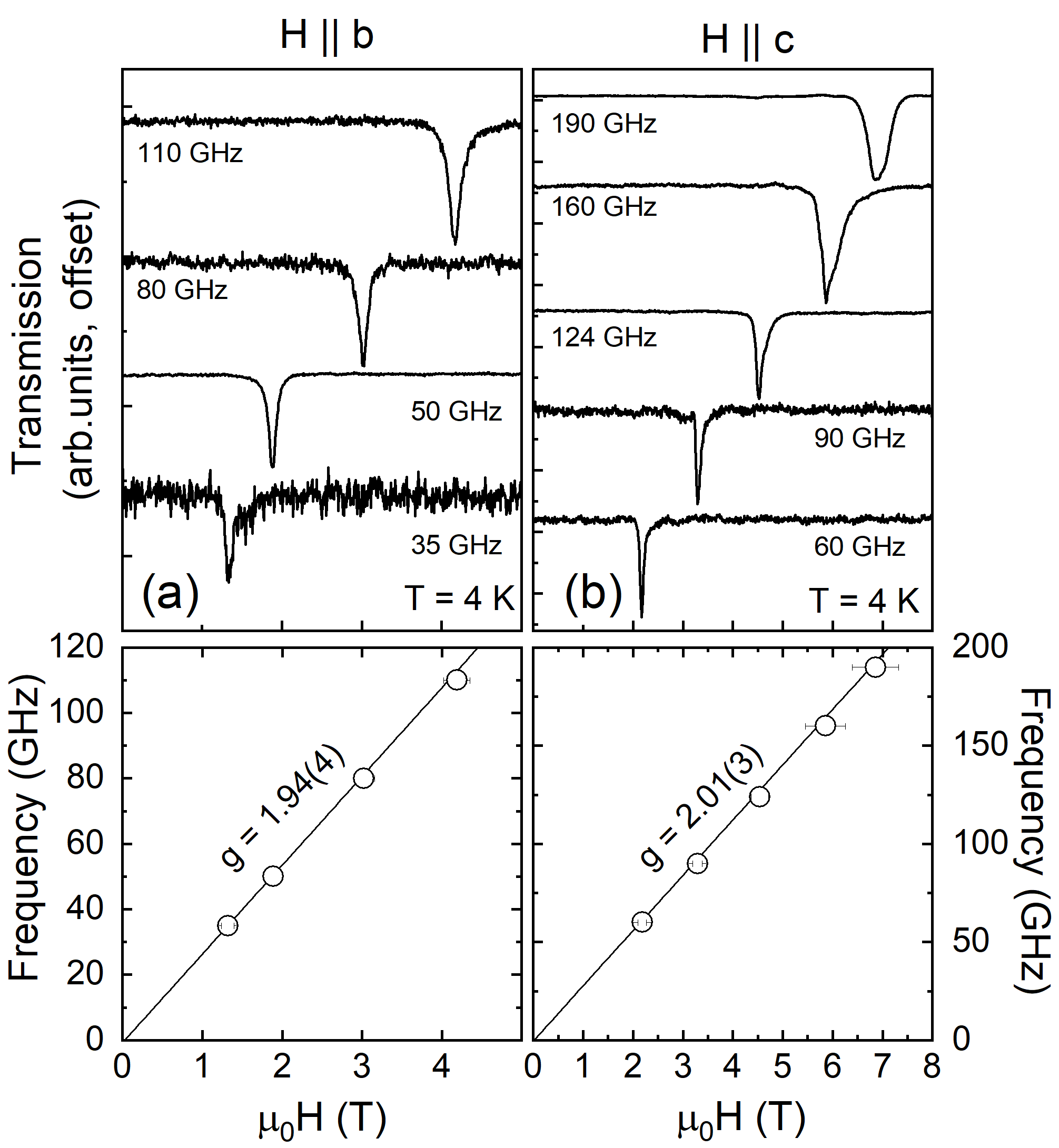}
\caption{Typical ESR spectra of Ca$_{10}$Cr$_7$O$_{28}$ at various frequencies and at $T=4$~K for (a) $H||b$  and (b) $H||c$, and corresponding resonance frequency-magnetic field diagrams.}
\end{figure}

Figure 2 displays the frequency dependence of the HF-ESR spectra measured at $T=4$~K at frequencies 35~GHz~$\leq \nu \leq$~190~GHz. The external magnetic field is applied along the $b$- and $c$-axis, respectively. For frequencies below 110~GHz, the ESR spectra consist of a single Lorentzian absorption line which originates from paramagnetic Cr$^{5+}$($3d^1$)-ions. The spectra are exchange-narrowed due to fast electronic fluctuations induced by exchange interaction between the Cr$^{5+}$-spins. For frequencies higher than 110~GHz and for magnetic fields $H||c$ applied perpendicular to the layers, the resonance line becomes asymmetric featuring a shoulder. The origin of the shoulder is not completely clear. In a spin-frustrated system like a kagome lattice, spin-spin correlations would induce internal fields around the magnetic ions, resulting in an additional feature deviating slightly from the main resonance position in the ESR spectra~\cite{Wellm}. Another possible reason for the shoulder feature is the mixing of phase and amplitude signals, which thus may hinder appropriate phase correction of the resonance feature. We also note that the (sub-)mm-sized samples at certain frequencies yielded multiple reflection features. Note, that possible variation of $g$-values due to different Cr$^{5+}$ sites yielding such a shoulder feature would imply, at 9.12~GHz, line splitting of $\geq 100$~mT which can be excluded by the X-band data. Thus, HF-ESR cannot reliably track the linewidth, so our analysis instead utilizes the X-band data for this purpose. However, the main resonance feature in our spectra is clear and sharp enough to be precisely analyzed and the contribution of the shoulder feature is relatively weak. The main peak is therefore used to construct the resonance frequency $vs.$ magnetic field diagrams for $H||b$ and $H||c$, as shown in Fig.~2. The resonance fields show linear behavior with vanishing zero-field energy gap. The obtained $g$-values, $g_b = 1.94(4)$ and $g_c =2.01(3)$, are typical for the high valence state of tetrahedrally coordinated Cr$^{5+}$ ions~\cite{Cage}.

\begin{figure}
\label{figure3}
\includegraphics[width=8.5cm]{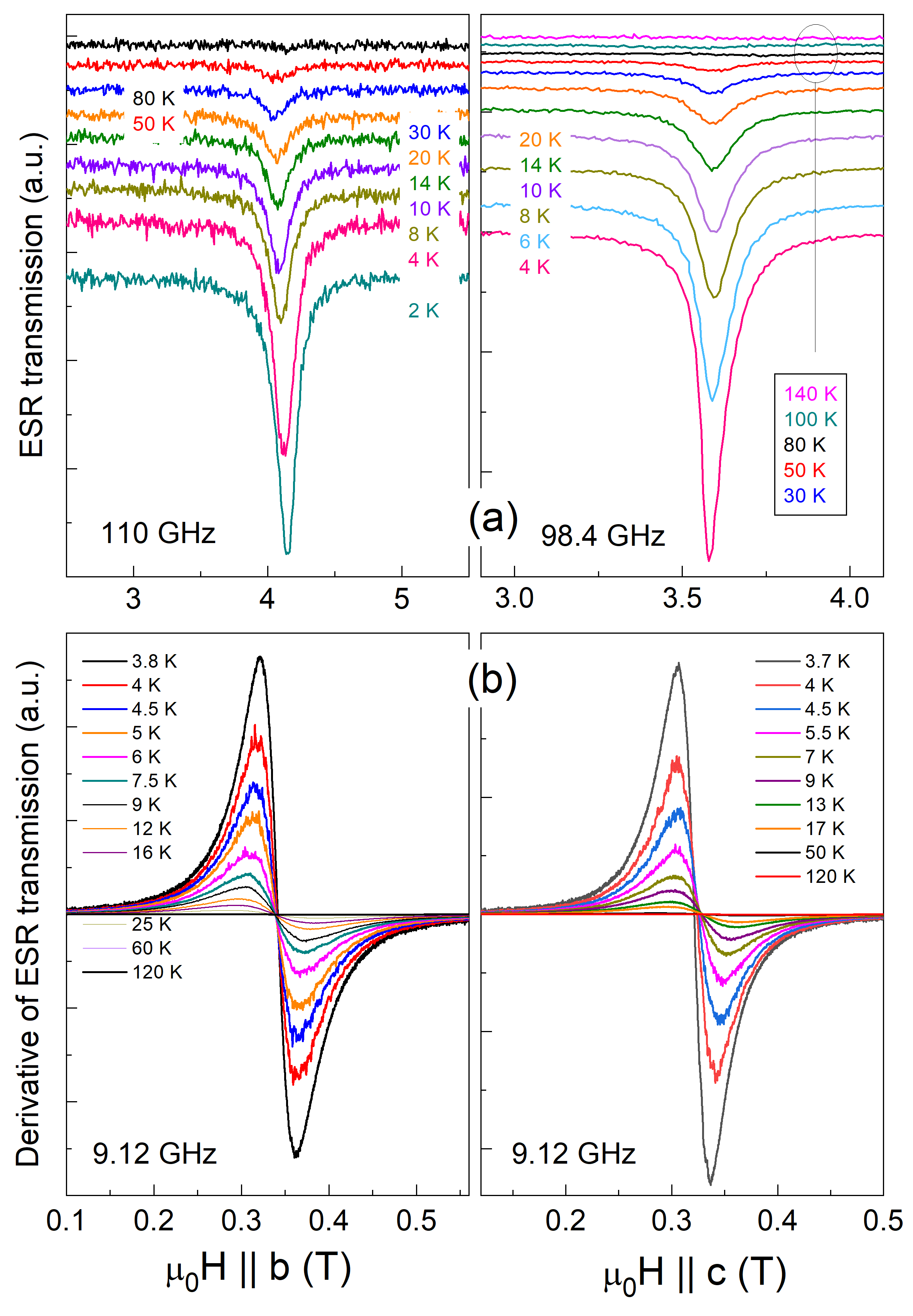}
\caption{Temperature dependence of the ESR spectra measured (a) at $\nu = 110$~GHz and $\nu = 98.4$~GHz for $H||b$ and $H||c$ on the left and right-hand sides respectively, and (b) at $\nu = 9.12$~GHz for $H||b$ and $H||c$ on the left and right-hand sides respectively. The Spectra in (a) are offset for clarity. \label{fig3}}
\end{figure}

Figure 3 shows the temperature dependence of the ESR signal recorded at $\nu=110$ and 9.12~GHz for $H||b$ as well as at $\nu=98.4$ and 9.12~GHz for $H||c$. For temperatures below 80~K, the data show the development of a substantial signal whose intensity increases upon cooling. For a quantitative analysis, the ESR spectra are fitted by a single Lorentzian profile. The ESR integrated intensity, the effective $g$-factor $g=h\nu /\mu_{\rm B}\mu_0 H_{\rm res}$ (with $H_{\rm res}$ being the resonance field), and the linewidth $\mu_0 \Delta H$ extracted from the fitting are plotted as a function of temperature in Figs.~4, 5 and 6, respectively.

\begin{figure}
\label{figure4}
\includegraphics[width=8.5cm]{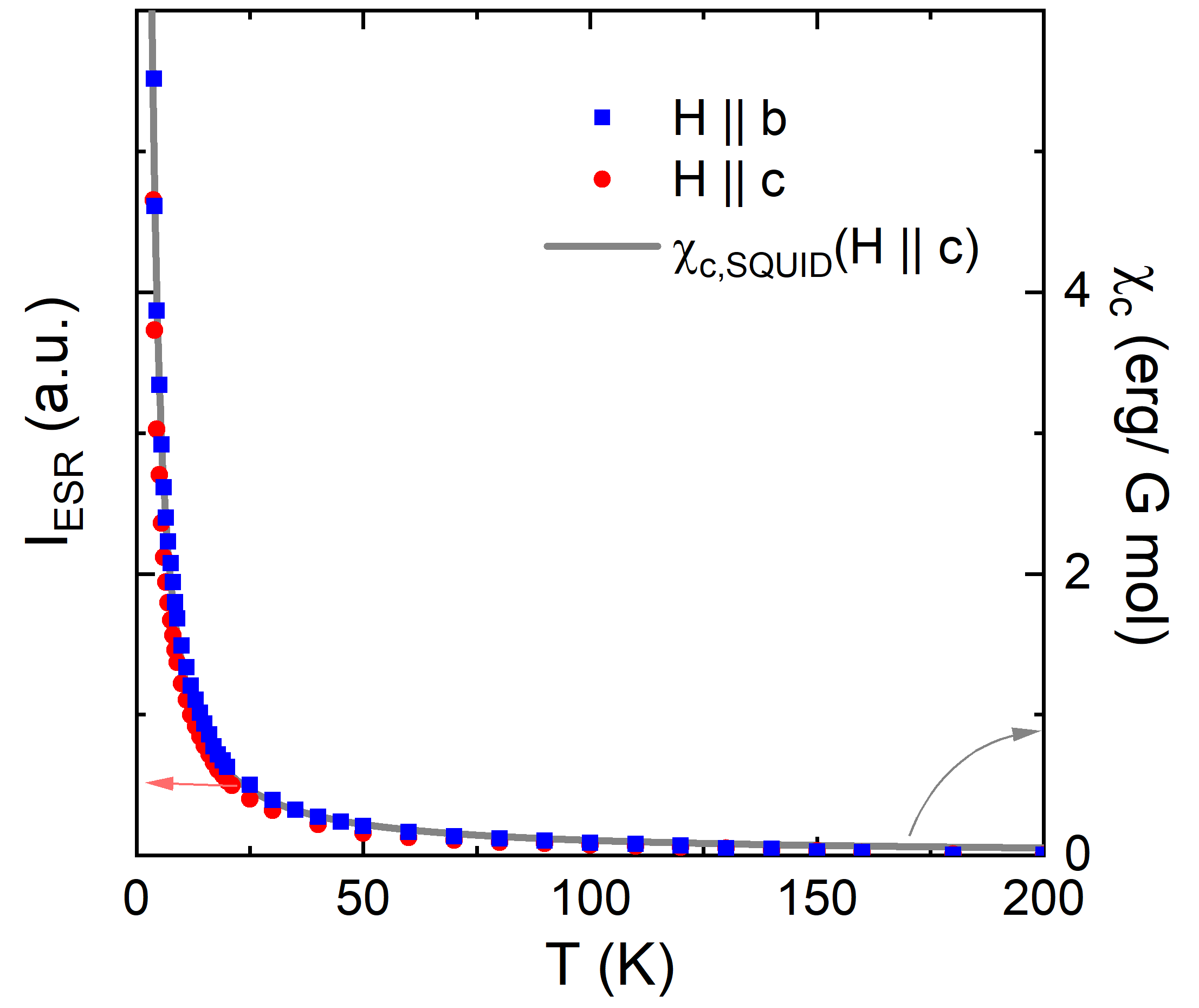}
\caption{Temperature dependence of the X-band ESR integrated intensity measured for $H||c$ and $H||b$, respectively. The grey line represents the static magnetic susceptibility $\chi_c$~\cite{Balz3}.\label{fig4}}
\end{figure}

The integrated intensity of the ESR resonance, $I_{\rm ESR}$, probes the intrinsic susceptibility $\chi_{\rm ESR}$ of the associated spins. For both magnetic field directions, $\chi_{\rm ESR}$ steeply increases with decreasing temperature (see Fig.~\ref{fig4}). In particular, the observed steady increase of $\chi_{\rm ESR}$ without saturation or reaching a peak exhibits no indication of a magnetic phase transition in the measured temperature range. Also, any feature associated with a gap in the excitation spectrum is absent. Typically, such a gap would cause a decrease in $\chi_{\rm ESR}$ at low temperatures due to the depletion of thermally populated triplet states. In agreement with previous studies~\cite{Sonnenschein,Ni}, this either implies a gapless ground state or that the spin–gap onset temperature is much smaller than the measured temperature range. We further note that the ESR integrated intensity agrees very well with the static magnetic susceptibility (Fig.~\ref{fig4}), suggesting that the ESR measurement and the static measurements are probing the same spins.

\begin{figure}
\label{figure5}
\includegraphics[width=8.5cm]{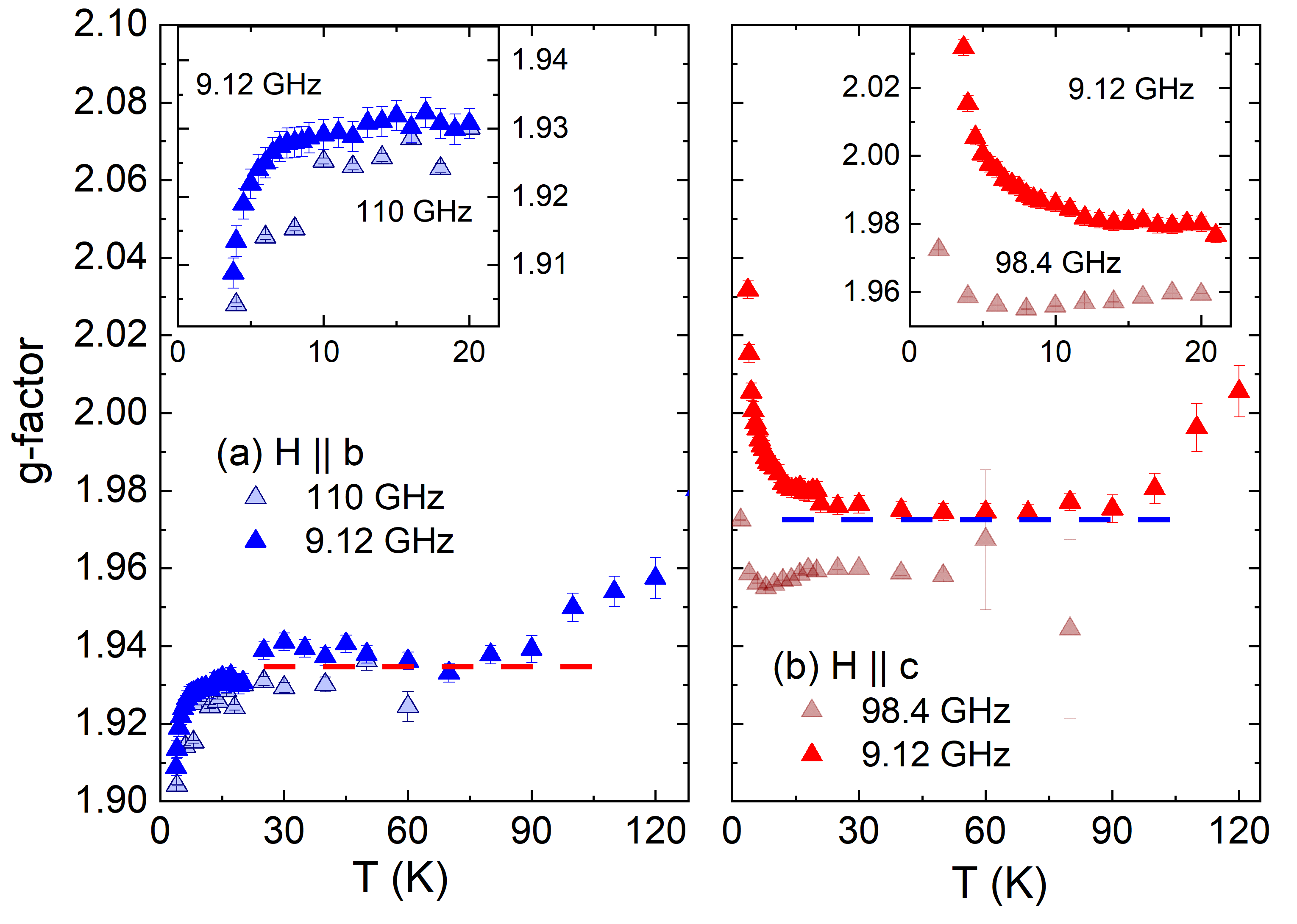}
\caption{Temperature dependence of the effective $g$-factor $g=h\nu /\mu_{\rm B}\mu_0 H_{\rm res}$ (a) at $\nu = 9.12$~GHz and $\nu = 110$~GHz for $H||b$, and (b) at $\nu = 9.12$~GHz and $\nu = 98.4$~GHz for $H||c$. Dashed lines are constant guide lines. Insets highlight the behavior below 20~K.\label{fig5}}
\end{figure}

The temperature dependence of the effective $g$-factor and hence of the resonance fields $H_{\rm res}$ implies different temperature regimes, as exhibited in Fig.~\ref{fig5}. As the temperature increases through 90~K, the $g$-factor increases for both $H||c$ and $H||b$, while it remains temperature-independent in the intermediate range between 20 and 90~K. In the low-temperature region, the X-band $g$-values shift downward for $H||b$ and upward for $H||c$ with decreasing temperature below 20~K. Noteworthy is that the $g$-factor obtained from the 98.4~GHz data initially decreases, reaching a minimum around 9~K before undergoing an upturn.

The observed shift of the $g$-factors signals the evolution of internal magnetic fields at the paramagnetic sites below $\approx 20$~K, thereby indicating the evolution of magnetic correlations in this temperature regime~\cite{Nagata1972,Yamada}. The upturn of the $g$-factor for $H||c$ indicates that the internal field along the $c-$axis is weaker than the internal field along $H||b$, where the downturn of the $g-$factor occurs, representing a stronger internal field induced by the strong intralayer interactions within the plane, also supporting the finding of 2D magnetism. This agrees with the observation of magnetic entropy changes in the same temperature regime~\cite{Balz2}. The $g$-values obtained at 98.4~GHz for $H||c$ which reflect the behavior at the resonance field of $\sim 3.6$~T significantly differ from the behavior in the X-band data, i.e., at $\sim 0.3$~T (Fig.~\ref{fig5}(b)). Associating the $g$-shift with the internal field implies strong field-induced changes of spin correlations for $H||c$. The nearly $T-$independent $g-$factor between 20 and 90~K is typical for a simple paramagnetic state, while the increasing $g-$factor above 90~K alludes to the presence of additional relaxation channels. No structural changes have been reported in the temperature range from 300 K down to 2 K~\cite{Balz3}. 

\begin{figure}
\label{figure6}
\includegraphics[width=8.5cm]{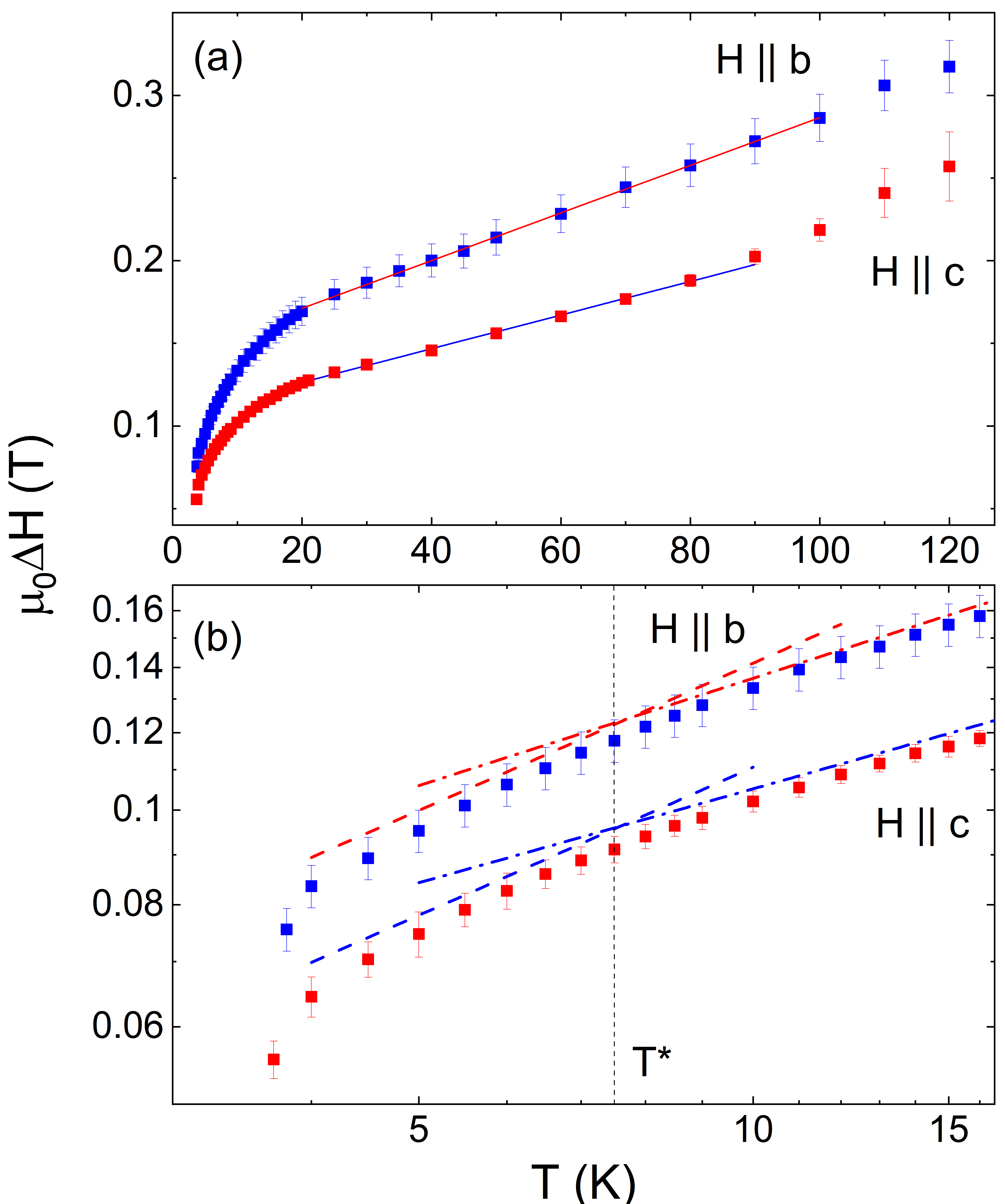}
\caption{(a) Temperature dependence of the X-band ESR linewidth for $H||b$ and $H||c$, respectively. Solid lines represent the linear fitting in the temperature range from 20~K to about 100~K.  (b) ESR linewidth at low temperatures below 16~K in a log-log plot. Vertical dash line indicates $T^*\approx 7.5$~K as discussed in the text. Dash lines and dash-dot lines are guides for eye below and above $T^*$, respectively.\label{fig6}}
\end{figure}

Similar to the evolution of internal fields, distinct regimes are also discernible in the temperature dependence of the EPR linewidth $\mu_0\Delta H$, as shown in Fig.~\ref{fig6}(a). Above about 20~K, a continuous increase of the linewidth is observed upon heating for both magnetic field directions. In the restricted temperature range, the linear behavior of the linewidth is confirmed with the linear fitting indicated as the solid lines in Fig.~\ref{fig6}(a). Above 100~K, the temperature dependence of the linewidth exhibits further changes in particular for $H||c$ (see Fig.~6(a)). In the high-$T$ paramagnetic (exchange-narrowing) regime, $\mu_0\Delta H$ is usually temperature independent as spin correlations are dying out above the characteristic temperature, $J/k_{\rm B}$, and a (modified) Kubo-Tomita limit may be reached~\cite{Kubo,Oshikawa,Choukroun}. Thus, the pronounced high-$T$ increase of $\mu_0\Delta H$ indicates the presence of an additional relaxation mechanism with changing processes through $90-100$~K. The linewidth, at 20~K, is found to be $\sim 0.12-0.17$~T, and increases to nearly 0.3~T ($H||b$) at 120~K. These values by far exceed the linewidths in Cr$^{5+}$-systems exhibiting only pure spin-spin relaxation~\cite{Dalal1981,Cage}. We attribute the observed behaviour of the linewidth to a dominant relaxation process via phonons~\cite{Huber,Heinrich,Eremin2008,Werner2016}. 

Upon cooling, $\mu_0\Delta H$ exhibits a change in temperature dependence at around 20~K signalling additional sharpening of the ESR line (see Fig.~\ref{fig6}(a)). As will be discussed below, the data also imply a crossover regime below $\sim 7.5$~K (see Fig.~\ref{fig6}(b)). The observed persistent decrease of the linewidth with decreasing temperature below $\sim 20$~K is, in particular, not typical for two-dimensional kagome systems, which are supposed to evolve short-range spin correlations below $k_{\rm B}T\sim J$ leading to an increase of linewidth upon cooling~\cite{Zorko,Zhang2010,Zorko2013}. The absence of critical line broadening down to the lowest measured temperature $T=3.6$~K excludes the presence of critical fluctuations of the staggered magnetization and is thus consistent with the lack of static magnetic order. Due to the restricted temperature regime where the smooth linewidth decreasing is observed, a quantitative estimate can, however, not reliably be performed.

We note that the low-temperature regime of linewidth decrease persists up to the effective magnetic energy scale ($J_{\rm eff}$/$k_{\rm B}$) of about 16 to 20~K, as inferred from (i) the saturation field of magnetization and specific heat of 12~T indicating magnetic interactions of about 1.4~meV~\cite{Balz2}, (ii) the deviation of the magnetic susceptibility from the Curie-Weiss behavior below 20~K~\cite{Balz2,Balodhi}, and (iii) the upper temperature of the broad magnetic specific heat peak suggesting magnetic entropy changes only below 20~K with spin correlations persisting up to this temperature~\cite{Balz2,Balodhi}. We associate the power-law decrease of the linewidth to the intrinsic spin dynamics of the alternating bilayer kagome system sensitively probed by the ESR linewidth. We emphasize that phonon-assisted relaxation is negligible and no structural instabilities are present at low temperatures. Instead, the power-law dependence of the ESR linewidth is often related to spin correlation dynamics in magnetic systems.

The smoothly decreasing linewidth below 16~K (i.e., where $T\lesssim J_{\mathrm{eff}}$) displays two distinct regimes separated around $T^* \approx 7.5$~K. In order to clarify these regimes, the linewidth below 16~K is plotted in a log-log scale in Fig.~\ref{fig6}(b). The difference between the two regimes is highlighted with the two power-law guide lines, $\mu_0\Delta H  \propto T^n$, where the exponent n conveys information about the underlying nature of spin excitations. Below $T^*$, the exponent $n$ is 0.51(8) and 0.52(6), and above $T^*$, $n$ is 0.37(8) and 0.33(3), for $H||b$ and $H||c$, respectively. This change in $n$ suggests an alteration in the underlying magnetic correlations, which evolves smoothly over temperature, as indicated in Ref. 45~\cite{Choi2012}. The two distinct regimes separated around $T^*$ are characterized by a change of anisotropy and accordingly by a change in spin excitations which suggests the presence of two magnetic energy scales. It is noteworthy that the crossover temperature $T^*/k_{\rm B}\simeq 0.6$~meV detected by the ESR linewidth agrees well with the energy separating two distinct spin excitation bands observed in INS, the lower one ranging from 0-0.6~meV and the upper one covering 0.7-1.5~meV~\cite{Balz1,Balz2}. Further, temperature-dependent muon-spin relaxation rates and stretching exponents obtained from $\mu$SR data exhibit an increase below $\sim7$~K~\cite{Balz1}. In this respect it may reflect the two different types of correlations at two distinct time-scales, as numerically identified by Pohle $et~al.$~\cite{Pohle}. More specifically, the weaker power-law dependence of $\mu_0\Delta H$ is attributed to magnetic correlations in the effective $S=3/2$ honeycomb lattice, whereas the stronger power-law dependence reflects a crossover to spin dynamics pertinent to the $S=1/2$ kagome antiferromagnet. Taken together, the two-stage behaviors observed in INS and ESR measurements suggest that the studied compound is effectively described by two distinct magnetic sublattices emerging from the bilayer kagome lattice, where exchange interactions feature opposing signs and different energy scales.

\section{Conclusions}

In summary, we have presented an X-band and HF-ESR study of the spiral spin liquid candidate material Ca$_{10}$Cr$_7$O$_{28}$. At higher temperatures, the temperature dependence of the ESR linewidth suggests the presence of an additional, presumingly phonon-related relaxation process. Upon cooling, the linewidth and the resonance field positions imply the evolution of two-stage spin  correlations below $\sim 20$~K. In the correlated paramagnetic regime ($k_{\rm B}T<J_{\mathrm{eff}}$), the linewidth exhibits power-law narrowing with decreasing temperature, not typical for two-dimensional uniform kagome systems. Notably, the ESR data reveal two distinct regimes of power-law behavior of the linewidth, with a crossover at $T^*\approx 7.5$~K, which are interpreted as the emergence of two magnetic sublattices with disparate magnetic correlations. This observation highlights that frustrated spin systems with complex exchange interactions of opposite signs and varying strengths can effectively be mapped onto subsystems with simpler spin topologies.

\begin{center}
{\bf ACKNOWLEDGEMENTS}
\end{center}

This work was supported by Deutsche Forschungsgemeinschaft (DFG) under Germany's Excellence Strategy EXC2181/1-390900948 (the Heidelberg STRUCTURES Excellence Cluster) and by Korea NRF Grant (No.~2020R1A2C3012367). Partial support by the DFG within project B06 of the SFB 1143 (Project No.~247310070) and by BMBF via the SpinFun project (13XP5088) is gratefully acknowledged. C.K. acknowledges support from the DFG (Project No.~415157846). S.L. acknowledges support from the Institute for Basic Science (IBS-R014-Y2).

\bibliography{CaCrO-Koo}

\begin{thebibliography}{45}%
\makeatletter
\providecommand \@ifxundefined [1]{%
 \@ifx{#1\undefined}
}%
\providecommand \@ifnum [1]{%
 \ifnum #1\expandafter \@firstoftwo
 \else \expandafter \@secondoftwo
 \fi
}%
\providecommand \@ifx [1]{%
 \ifx #1\expandafter \@firstoftwo
 \else \expandafter \@secondoftwo
 \fi
}%
\providecommand \natexlab [1]{#1}%
\providecommand \enquote  [1]{``#1''}%
\providecommand \bibnamefont  [1]{#1}%
\providecommand \bibfnamefont [1]{#1}%
\providecommand \citenamefont [1]{#1}%
\providecommand \href@noop [0]{\@secondoftwo}%
\providecommand \href [0]{\begingroup \@sanitize@url \@href}%
\providecommand \@href[1]{\@@startlink{#1}\@@href}%
\providecommand \@@href[1]{\endgroup#1\@@endlink}%
\providecommand \@sanitize@url [0]{\catcode `\\12\catcode `\$12\catcode `\&12\catcode `\#12\catcode `\^12\catcode `\_12\catcode `\%12\relax}%
\providecommand \@@startlink[1]{}%
\providecommand \@@endlink[0]{}%
\providecommand \url  [0]{\begingroup\@sanitize@url \@url }%
\providecommand \@url [1]{\endgroup\@href {#1}{\urlprefix }}%
\providecommand \urlprefix  [0]{URL }%
\providecommand \Eprint [0]{\href }%
\providecommand \doibase [0]{http://dx.doi.org/}%
\providecommand \selectlanguage [0]{\@gobble}%
\providecommand \bibinfo  [0]{\@secondoftwo}%
\providecommand \bibfield  [0]{\@secondoftwo}%
\providecommand \translation [1]{[#1]}%
\providecommand \BibitemOpen [0]{}%
\providecommand \bibitemStop [0]{}%
\providecommand \bibitemNoStop [0]{.\EOS\space}%
\providecommand \EOS [0]{\spacefactor3000\relax}%
\providecommand \BibitemShut  [1]{\csname bibitem#1\endcsname}%
\let\auto@bib@innerbib\@empty
\bibitem [{\citenamefont {Balents}(2010)}]{Balents}%
  \BibitemOpen
  \bibfield  {author} {\bibinfo {author} {\bibfnamefont {L.}~\bibnamefont {Balents}},\ }\bibfield  {title} {\enquote {\bibinfo {title} {Spin liquids in frustrated magnets},}\ }\href {https://www.nature.com/articles/nature08917} {\bibfield  {journal} {\bibinfo  {journal} {Nature}\ }\textbf {\bibinfo {volume} {464}},\ \bibinfo {pages} {199} (\bibinfo {year} {2010})}\BibitemShut {NoStop}%
\bibitem [{\citenamefont {Wen}(1991)}]{Wen}%
  \BibitemOpen
  \bibfield  {author} {\bibinfo {author} {\bibfnamefont {X.~G.}\ \bibnamefont {Wen}},\ }\bibfield  {title} {\enquote {\bibinfo {title} {Mean-field theory of spin-liquid states with finite energy gap and topological orders},}\ }\href {https://journals.aps.org/prb/abstract/10.1103/PhysRevB.44.2664} {\bibfield  {journal} {\bibinfo  {journal} {Phys. Rev. B}\ }\textbf {\bibinfo {volume} {44}},\ \bibinfo {pages} {2664} (\bibinfo {year} {1991})}\BibitemShut {NoStop}%
\bibitem [{\citenamefont {Lake}\ \emph {et~al.}(2005)\citenamefont {Lake}, \citenamefont {Tennant}, \citenamefont {Frost},\ and\ \citenamefont {Nagler}}]{Lake}%
  \BibitemOpen
  \bibfield  {author} {\bibinfo {author} {\bibfnamefont {B.}~\bibnamefont {Lake}}, \bibinfo {author} {\bibfnamefont {D.}~\bibnamefont {Tennant}}, \bibinfo {author} {\bibfnamefont {C.}~\bibnamefont {Frost}}, \ and\ \bibinfo {author} {\bibfnamefont {S.}~\bibnamefont {Nagler}},\ }\bibfield  {title} {\enquote {\bibinfo {title} {Quantum criticality and universal scaling of a quantum antiferromagnet},}\ }\href {https://www.nature.com/articles/nmat1327} {\bibfield  {journal} {\bibinfo  {journal} {Nat. Mater.}\ }\textbf {\bibinfo {volume} {4}},\ \bibinfo {pages} {329} (\bibinfo {year} {2005})}\BibitemShut {NoStop}%
\bibitem [{\citenamefont {Mourigal}\ \emph {et~al.}(2013)\citenamefont {Mourigal}, \citenamefont {Enderle}, \citenamefont {Kloepperpieper}, \citenamefont {Caux}, \citenamefont {Stunault},\ and\ \citenamefont {R{\o}nnow}}]{Mourigal}%
  \BibitemOpen
  \bibfield  {author} {\bibinfo {author} {\bibfnamefont {M.}~\bibnamefont {Mourigal}}, \bibinfo {author} {\bibfnamefont {M.}~\bibnamefont {Enderle}}, \bibinfo {author} {\bibfnamefont {A.}~\bibnamefont {Kloepperpieper}}, \bibinfo {author} {\bibfnamefont {J.-S.}\ \bibnamefont {Caux}}, \bibinfo {author} {\bibfnamefont {A.}~\bibnamefont {Stunault}}, \ and\ \bibinfo {author} {\bibfnamefont {H.~M.}\ \bibnamefont {R{\o}nnow}},\ }\bibfield  {title} {\enquote {\bibinfo {title} {Fractional spinon excitations in the quantum heisenberg antiferromagnetic chain},}\ }\href {https://www.nature.com/articles/nphys2652} {\bibfield  {journal} {\bibinfo  {journal} {Nat. Phys.}\ }\textbf {\bibinfo {volume} {9}},\ \bibinfo {pages} {435} (\bibinfo {year} {2013})}\BibitemShut {NoStop}%
\bibitem [{\citenamefont {Broholm}\ \emph {et~al.}(2020)\citenamefont {Broholm}, \citenamefont {Cava}, \citenamefont {Kivelson}, \citenamefont {Nocera}, \citenamefont {Norman},\ and\ \citenamefont {Senthil}}]{BroholmScience2020}%
  \BibitemOpen
  \bibfield  {author} {\bibinfo {author} {\bibfnamefont {C.}~\bibnamefont {Broholm}}, \bibinfo {author} {\bibfnamefont {R.~J.}\ \bibnamefont {Cava}}, \bibinfo {author} {\bibfnamefont {S.~A.}\ \bibnamefont {Kivelson}}, \bibinfo {author} {\bibfnamefont {D.~G.}\ \bibnamefont {Nocera}}, \bibinfo {author} {\bibfnamefont {M.~R.}\ \bibnamefont {Norman}}, \ and\ \bibinfo {author} {\bibfnamefont {T.}~\bibnamefont {Senthil}},\ }\bibfield  {title} {\enquote {\bibinfo {title} {Quantum spin liquids},}\ }\href {https://www.science.org/doi/10.1126/science.aay0668} {\bibfield  {journal} {\bibinfo  {journal} {Science}\ }\textbf {\bibinfo {volume} {367}},\ \bibinfo {pages} {263} (\bibinfo {year} {2020})}\BibitemShut {NoStop}%
\bibitem [{\citenamefont {Shimizu}\ \emph {et~al.}(2003)\citenamefont {Shimizu}, \citenamefont {Miyagawa}, \citenamefont {Kanoda}, \citenamefont {Maesato},\ and\ \citenamefont {Saito}}]{Shimizu}%
  \BibitemOpen
  \bibfield  {author} {\bibinfo {author} {\bibfnamefont {Y.}~\bibnamefont {Shimizu}}, \bibinfo {author} {\bibfnamefont {K.}~\bibnamefont {Miyagawa}}, \bibinfo {author} {\bibfnamefont {K.}~\bibnamefont {Kanoda}}, \bibinfo {author} {\bibfnamefont {M.}~\bibnamefont {Maesato}}, \ and\ \bibinfo {author} {\bibfnamefont {G.}~\bibnamefont {Saito}},\ }\bibfield  {title} {\enquote {\bibinfo {title} {Spin liquid state in an organic mott insulator with a triangular lattice},}\ }\href {https://journals.aps.org/prl/abstract/10.1103/PhysRevLett.91.107001} {\bibfield  {journal} {\bibinfo  {journal} {Phys. Rev. Lett.}\ }\textbf {\bibinfo {volume} {91}},\ \bibinfo {pages} {107001} (\bibinfo {year} {2003})}\BibitemShut {NoStop}%
\bibitem [{\citenamefont {Shen}\ \emph {et~al.}(2016)\citenamefont {Shen}, \citenamefont {Li}, \citenamefont {Wo}, \citenamefont {Li}, \citenamefont {Shen}, \citenamefont {Pan}, \citenamefont {Wang}, \citenamefont {Walker}, \citenamefont {Steffens}, \citenamefont {Boehm}, \citenamefont {Hao}, \citenamefont {Quintero-Castro}, \citenamefont {Harriger}, \citenamefont {Frontzek}, \citenamefont {Hao}, \citenamefont {Meng}, \citenamefont {Zhang}, \citenamefont {Chen},\ and\ \citenamefont {Zhao}}]{Shen}%
  \BibitemOpen
  \bibfield  {author} {\bibinfo {author} {\bibfnamefont {Y.}~\bibnamefont {Shen}}, \bibinfo {author} {\bibfnamefont {Y.-D.}\ \bibnamefont {Li}}, \bibinfo {author} {\bibfnamefont {H.}~\bibnamefont {Wo}}, \bibinfo {author} {\bibfnamefont {Y.}~\bibnamefont {Li}}, \bibinfo {author} {\bibfnamefont {S.}~\bibnamefont {Shen}}, \bibinfo {author} {\bibfnamefont {B.}~\bibnamefont {Pan}}, \bibinfo {author} {\bibfnamefont {Q.}~\bibnamefont {Wang}}, \bibinfo {author} {\bibfnamefont {H.~C.}\ \bibnamefont {Walker}}, \bibinfo {author} {\bibfnamefont {P.}~\bibnamefont {Steffens}}, \bibinfo {author} {\bibfnamefont {M.}~\bibnamefont {Boehm}}, \bibinfo {author} {\bibfnamefont {Y.}~\bibnamefont {Hao}}, \bibinfo {author} {\bibfnamefont {D.~L.}\ \bibnamefont {Quintero-Castro}}, \bibinfo {author} {\bibfnamefont {L.~W.}\ \bibnamefont {Harriger}}, \bibinfo {author} {\bibfnamefont {M.~D.}\ \bibnamefont {Frontzek}}, \bibinfo {author} {\bibfnamefont {L.}~\bibnamefont {Hao}}, \bibinfo {author} {\bibfnamefont {S.}~\bibnamefont {Meng}},
  \bibinfo {author} {\bibfnamefont {Q.}~\bibnamefont {Zhang}}, \bibinfo {author} {\bibfnamefont {G.}~\bibnamefont {Chen}}, \ and\ \bibinfo {author} {\bibfnamefont {J.}~\bibnamefont {Zhao}},\ }\bibfield  {title} {\enquote {\bibinfo {title} {Evidence for a spinon fermi surface in a triangular-lattice quantum-spin-liquid candidate},}\ }\href {https://www.nature.com/articles/nature20614} {\bibfield  {journal} {\bibinfo  {journal} {Nature}\ }\textbf {\bibinfo {volume} {540}},\ \bibinfo {pages} {559} (\bibinfo {year} {2016})}\BibitemShut {NoStop}%
\bibitem [{\citenamefont {Pratt}\ \emph {et~al.}(2011)\citenamefont {Pratt}, \citenamefont {Baker}, \citenamefont {Blundell}, \citenamefont {Lancaster}, \citenamefont {Ohira-Kawamura}, \citenamefont {Baines}, \citenamefont {Shimizu}, \citenamefont {Kanoda}, \citenamefont {Watanabe},\ and\ \citenamefont {Saito}}]{Pratt}%
  \BibitemOpen
  \bibfield  {author} {\bibinfo {author} {\bibfnamefont {F.~L.}\ \bibnamefont {Pratt}}, \bibinfo {author} {\bibfnamefont {P.~J.}\ \bibnamefont {Baker}}, \bibinfo {author} {\bibfnamefont {S.~J.}\ \bibnamefont {Blundell}}, \bibinfo {author} {\bibfnamefont {T.}~\bibnamefont {Lancaster}}, \bibinfo {author} {\bibfnamefont {S.}~\bibnamefont {Ohira-Kawamura}}, \bibinfo {author} {\bibfnamefont {C.}~\bibnamefont {Baines}}, \bibinfo {author} {\bibfnamefont {Y.}~\bibnamefont {Shimizu}}, \bibinfo {author} {\bibfnamefont {K.}~\bibnamefont {Kanoda}}, \bibinfo {author} {\bibfnamefont {I.}~\bibnamefont {Watanabe}}, \ and\ \bibinfo {author} {\bibfnamefont {G.}~\bibnamefont {Saito}},\ }\bibfield  {title} {\enquote {\bibinfo {title} {Magnetic and non-magnetic phases of a quantum spin liquid},}\ }\href {https://www.nature.com/articles/nature09910} {\bibfield  {journal} {\bibinfo  {journal} {Nature}\ }\textbf {\bibinfo {volume} {471}},\ \bibinfo {pages} {612} (\bibinfo {year} {2011})}\BibitemShut {NoStop}%
\bibitem [{\citenamefont {Lee}\ and\ \citenamefont {Lee}(2005)}]{Lee}%
  \BibitemOpen
  \bibfield  {author} {\bibinfo {author} {\bibfnamefont {S.-S.}\ \bibnamefont {Lee}}\ and\ \bibinfo {author} {\bibfnamefont {P.~A.}\ \bibnamefont {Lee}},\ }\bibfield  {title} {\enquote {\bibinfo {title} {U(1) gauge theory of the hubbard model: Spin liquid states and possible application to $\kappa$-{(BEDT-TTF)}$_2${⁢Cu}$_2${⁢(CN)}$_3$},}\ }\href {https://journals.aps.org/prl/abstract/10.1103/PhysRevLett.95.036403} {\bibfield  {journal} {\bibinfo  {journal} {Phys. Rev. Lett.}\ }\textbf {\bibinfo {volume} {95}},\ \bibinfo {pages} {036403} (\bibinfo {year} {2005})}\BibitemShut {NoStop}%
\bibitem [{\citenamefont {Yan}\ \emph {et~al.}(2011)\citenamefont {Yan}, \citenamefont {Huse},\ and\ \citenamefont {White}}]{Yan}%
  \BibitemOpen
  \bibfield  {author} {\bibinfo {author} {\bibfnamefont {S.}~\bibnamefont {Yan}}, \bibinfo {author} {\bibfnamefont {D.~A.}\ \bibnamefont {Huse}}, \ and\ \bibinfo {author} {\bibfnamefont {S.~R.}\ \bibnamefont {White}},\ }\bibfield  {title} {\enquote {\bibinfo {title} {Spin-liquid ground state of the {S=1/2} kagome heisenberg antiferromagnet},}\ }\href {https://www.science.org/doi/10.1126/science.1201080} {\bibfield  {journal} {\bibinfo  {journal} {Science}\ }\textbf {\bibinfo {volume} {332}},\ \bibinfo {pages} {1173} (\bibinfo {year} {2011})}\BibitemShut {NoStop}%
\bibitem [{\citenamefont {Han}\ \emph {et~al.}(2012)\citenamefont {Han}, \citenamefont {Helton}, \citenamefont {Chu}, \citenamefont {Nocera}, \citenamefont {Rodriguez-Rivera}, \citenamefont {Broholm},\ and\ \citenamefont {Lee}}]{Han}%
  \BibitemOpen
  \bibfield  {author} {\bibinfo {author} {\bibfnamefont {T.-H.}\ \bibnamefont {Han}}, \bibinfo {author} {\bibfnamefont {J.~S.}\ \bibnamefont {Helton}}, \bibinfo {author} {\bibfnamefont {S.}~\bibnamefont {Chu}}, \bibinfo {author} {\bibfnamefont {D.~G.}\ \bibnamefont {Nocera}}, \bibinfo {author} {\bibfnamefont {J.~A.}\ \bibnamefont {Rodriguez-Rivera}}, \bibinfo {author} {\bibfnamefont {C.}~\bibnamefont {Broholm}}, \ and\ \bibinfo {author} {\bibfnamefont {Y.~S.}\ \bibnamefont {Lee}},\ }\bibfield  {title} {\enquote {\bibinfo {title} {Fractionalized excitations in the spin-liquid state of a kagome-lattice antiferromagnet},}\ }\href {https://www.nature.com/articles/nature11659} {\bibfield  {journal} {\bibinfo  {journal} {Nature}\ }\textbf {\bibinfo {volume} {492}},\ \bibinfo {pages} {406} (\bibinfo {year} {2012})}\BibitemShut {NoStop}%
\bibitem [{\citenamefont {Gingras}\ and\ \citenamefont {McClarty}(2014)}]{Gingras}%
  \BibitemOpen
  \bibfield  {author} {\bibinfo {author} {\bibfnamefont {M.~J.~P.}\ \bibnamefont {Gingras}}\ and\ \bibinfo {author} {\bibfnamefont {P.~A.}\ \bibnamefont {McClarty}},\ }\bibfield  {title} {\enquote {\bibinfo {title} {Quantum spin ice: a search for gapless quantum spin liquids in pyrochlore magnets},}\ }\href {https://iopscience.iop.org/article/10.1088/0034-4885/77/5/056501} {\bibfield  {journal} {\bibinfo  {journal} {Rep. Prog. Phys.}\ }\textbf {\bibinfo {volume} {77}},\ \bibinfo {pages} {056501} (\bibinfo {year} {2014})}\BibitemShut {NoStop}%
\bibitem [{\citenamefont {Li}\ \emph {et~al.}(2024)\citenamefont {Li}, \citenamefont {Rutherford}, \citenamefont {Wang}, \citenamefont {Liang}, \citenamefont {Li}, \citenamefont {Zhang}, \citenamefont {Wang}, \citenamefont {Xie}, \citenamefont {Zhou},\ and\ \citenamefont {Sun}}]{Li}%
  \BibitemOpen
  \bibfield  {author} {\bibinfo {author} {\bibfnamefont {N.}~\bibnamefont {Li}}, \bibinfo {author} {\bibfnamefont {A.}~\bibnamefont {Rutherford}}, \bibinfo {author} {\bibfnamefont {Y.~Y.}\ \bibnamefont {Wang}}, \bibinfo {author} {\bibfnamefont {H.}~\bibnamefont {Liang}}, \bibinfo {author} {\bibfnamefont {Q.~J.}\ \bibnamefont {Li}}, \bibinfo {author} {\bibfnamefont {Z.~J.}\ \bibnamefont {Zhang}}, \bibinfo {author} {\bibfnamefont {H.}~\bibnamefont {Wang}}, \bibinfo {author} {\bibfnamefont {W.}~\bibnamefont {Xie}}, \bibinfo {author} {\bibfnamefont {H.~D.}\ \bibnamefont {Zhou}}, \ and\ \bibinfo {author} {\bibfnamefont {X.~F.}\ \bibnamefont {Sun}},\ }\bibfield  {title} {\enquote {\bibinfo {title} {Ising-type quantum spin liquid state in {PrMgAl}$_{11}${O}$_{19}$},}\ }\href {https://journals.aps.org/prb/abstract/10.1103/PhysRevB.110.134401} {\bibfield  {journal} {\bibinfo  {journal} {Phys. Rev. B}\ }\textbf {\bibinfo {volume} {110}},\ \bibinfo {pages} {134401} (\bibinfo {year} {2024})}\BibitemShut {NoStop}%
\bibitem [{\citenamefont {Xu}\ \emph {et~al.}(2016)\citenamefont {Xu}, \citenamefont {Zhang}, \citenamefont {Li}, \citenamefont {Yu}, \citenamefont {Hong}, \citenamefont {Zhang},\ and\ \citenamefont {Li}}]{Xu}%
  \BibitemOpen
  \bibfield  {author} {\bibinfo {author} {\bibfnamefont {Y.}~\bibnamefont {Xu}}, \bibinfo {author} {\bibfnamefont {J.}~\bibnamefont {Zhang}}, \bibinfo {author} {\bibfnamefont {Y.~S.}\ \bibnamefont {Li}}, \bibinfo {author} {\bibfnamefont {Y.~J.}\ \bibnamefont {Yu}}, \bibinfo {author} {\bibfnamefont {X.~C.}\ \bibnamefont {Hong}}, \bibinfo {author} {\bibfnamefont {Q.~M.}\ \bibnamefont {Zhang}}, \ and\ \bibinfo {author} {\bibfnamefont {S.~Y.}\ \bibnamefont {Li}},\ }\bibfield  {title} {\enquote {\bibinfo {title} {Absence of magnetic thermal conductivity in the quantum spin-liquid candidate {YbMgGaO}$_4$},}\ }\href {https://journals.aps.org/prl/abstract/10.1103/PhysRevLett.117.267202} {\bibfield  {journal} {\bibinfo  {journal} {Phys. Rev. Lett.}\ }\textbf {\bibinfo {volume} {117}},\ \bibinfo {pages} {267202} (\bibinfo {year} {2016})}\BibitemShut {NoStop}%
\bibitem [{\citenamefont {Xu}\ \emph {et~al.}(2023)\citenamefont {Xu}, \citenamefont {Bag}, \citenamefont {Sherman}, \citenamefont {Yadav}, \citenamefont {Kolesnikov}, \citenamefont {Podlesnyak}, \citenamefont {Moore},\ and\ \citenamefont {Haravifard}}]{Xu23}%
  \BibitemOpen
  \bibfield  {author} {\bibinfo {author} {\bibfnamefont {S.}~\bibnamefont {Xu}}, \bibinfo {author} {\bibfnamefont {R.}~\bibnamefont {Bag}}, \bibinfo {author} {\bibfnamefont {N.~E.}\ \bibnamefont {Sherman}}, \bibinfo {author} {\bibfnamefont {L.}~\bibnamefont {Yadav}}, \bibinfo {author} {\bibfnamefont {A.~I.}\ \bibnamefont {Kolesnikov}}, \bibinfo {author} {\bibfnamefont {A.~A.}\ \bibnamefont {Podlesnyak}}, \bibinfo {author} {\bibfnamefont {J.~E.}\ \bibnamefont {Moore}}, \ and\ \bibinfo {author} {\bibfnamefont {S.}~\bibnamefont {Haravifard}},\ }\bibfield  {title} {\enquote {\bibinfo {title} {Realization of {U}(1) dirac quantum spin liquid in {YbZn}$_2${GaO}$_5$},}\ }\href {https://arxiv.org/abs/2305.20040} {\bibfield  {journal} {\bibinfo  {journal} {arXiv:2305.20040}\ } (\bibinfo {year} {2023})}\BibitemShut {NoStop}%
\bibitem [{\citenamefont {Kitaev}(2006)}]{Kitaev}%
  \BibitemOpen
  \bibfield  {author} {\bibinfo {author} {\bibfnamefont {A.}~\bibnamefont {Kitaev}},\ }\bibfield  {title} {\enquote {\bibinfo {title} {Anyons in an exactly solved model and beyond},}\ }\href {https://www.sciencedirect.com/science/article/pii/S0003491605002381} {\bibfield  {journal} {\bibinfo  {journal} {Ann. Phys.}\ }\textbf {\bibinfo {volume} {321}},\ \bibinfo {pages} {2} (\bibinfo {year} {2006})}\BibitemShut {NoStop}%
\bibitem [{\citenamefont {Baskaran}\ \emph {et~al.}(2007)\citenamefont {Baskaran}, \citenamefont {Mandal},\ and\ \citenamefont {Shankar}}]{Baskaran}%
  \BibitemOpen
  \bibfield  {author} {\bibinfo {author} {\bibfnamefont {G.}~\bibnamefont {Baskaran}}, \bibinfo {author} {\bibfnamefont {S.}~\bibnamefont {Mandal}}, \ and\ \bibinfo {author} {\bibfnamefont {R.}~\bibnamefont {Shankar}},\ }\bibfield  {title} {\enquote {\bibinfo {title} {Exact results for spin dynamics and fractionalization in the kitaev model},}\ }\href {https://journals.aps.org/prl/abstract/10.1103/PhysRevLett.98.247201} {\bibfield  {journal} {\bibinfo  {journal} {Phys. Rev. Lett.}\ }\textbf {\bibinfo {volume} {98}},\ \bibinfo {pages} {247201} (\bibinfo {year} {2007})}\BibitemShut {NoStop}%
\bibitem [{\citenamefont {Knolle}\ \emph {et~al.}(2014)\citenamefont {Knolle}, \citenamefont {Kovrizhin}, \citenamefont {Chalker},\ and\ \citenamefont {Moessner}}]{Knolle}%
  \BibitemOpen
  \bibfield  {author} {\bibinfo {author} {\bibfnamefont {J.}~\bibnamefont {Knolle}}, \bibinfo {author} {\bibfnamefont {D.~L.}\ \bibnamefont {Kovrizhin}}, \bibinfo {author} {\bibfnamefont {J.~T.}\ \bibnamefont {Chalker}}, \ and\ \bibinfo {author} {\bibfnamefont {R.}~\bibnamefont {Moessner}},\ }\bibfield  {title} {\enquote {\bibinfo {title} {Dynamics of a two-dimensional quantum spin liquid: Signatures of emergent majorana fermions and fluxes},}\ }\href {https://journals.aps.org/prl/abstract/10.1103/PhysRevLett.112.207203} {\bibfield  {journal} {\bibinfo  {journal} {Phys. Rev. Lett.}\ }\textbf {\bibinfo {volume} {112}},\ \bibinfo {pages} {207203} (\bibinfo {year} {2014})}\BibitemShut {NoStop}%
\bibitem [{\citenamefont {Balz}\ \emph {et~al.}(2016)\citenamefont {Balz}, \citenamefont {nad J.~Reuther}, \citenamefont {Luetkens}, \citenamefont {Sch{\o}nemann}, \citenamefont {Herrmannsd{\o}rfer}, \citenamefont {Singh}, \citenamefont {Islam}, \citenamefont {Wheeler}, \citenamefont {Rodriguez-Rivera}, \citenamefont {Guidi}, \citenamefont {Simeoni}, \citenamefont {Baines},\ and\ \citenamefont {Ryll}}]{Balz1}%
  \BibitemOpen
  \bibfield  {author} {\bibinfo {author} {\bibfnamefont {C.}~\bibnamefont {Balz}}, \bibinfo {author} {\bibfnamefont {B.~Lake}\ \bibnamefont {nad J.~Reuther}}, \bibinfo {author} {\bibfnamefont {H.}~\bibnamefont {Luetkens}}, \bibinfo {author} {\bibfnamefont {R.}~\bibnamefont {Sch{\o}nemann}}, \bibinfo {author} {\bibfnamefont {T.}~\bibnamefont {Herrmannsd{\o}rfer}}, \bibinfo {author} {\bibfnamefont {Y.}~\bibnamefont {Singh}}, \bibinfo {author} {\bibfnamefont {A.~T. M.~Nazmul}\ \bibnamefont {Islam}}, \bibinfo {author} {\bibfnamefont {E.~M.}\ \bibnamefont {Wheeler}}, \bibinfo {author} {\bibfnamefont {J.~A.}\ \bibnamefont {Rodriguez-Rivera}}, \bibinfo {author} {\bibfnamefont {T.}~\bibnamefont {Guidi}}, \bibinfo {author} {\bibfnamefont {G.~G.}\ \bibnamefont {Simeoni}}, \bibinfo {author} {\bibfnamefont {C.}~\bibnamefont {Baines}}, \ and\ \bibinfo {author} {\bibfnamefont {H.}~\bibnamefont {Ryll}},\ }\bibfield  {title} {\enquote {\bibinfo {title} {Physical realization of a quantum spin liquid based on a complex
  frustration mechanism},}\ }\href {https://www.nature.com/articles/nphys3826} {\bibfield  {journal} {\bibinfo  {journal} {Nat. Phys.}\ }\textbf {\bibinfo {volume} {12}},\ \bibinfo {pages} {942} (\bibinfo {year} {2016})}\BibitemShut {NoStop}%
\bibitem [{\citenamefont {Balz}\ \emph {et~al.}(2017{\natexlab{a}})\citenamefont {Balz}, \citenamefont {Lake}, \citenamefont {Islam}, \citenamefont {Singh}, \citenamefont {Rodriguez-Rivera}, \citenamefont {Guidi}, \citenamefont {Wheeler}, \citenamefont {Simeoni},\ and\ \citenamefont {Ryll}}]{Balz2}%
  \BibitemOpen
  \bibfield  {author} {\bibinfo {author} {\bibfnamefont {Ch.}\ \bibnamefont {Balz}}, \bibinfo {author} {\bibfnamefont {B.}~\bibnamefont {Lake}}, \bibinfo {author} {\bibfnamefont {A.T.M.~Nazmul}\ \bibnamefont {Islam}}, \bibinfo {author} {\bibfnamefont {Y.}~\bibnamefont {Singh}}, \bibinfo {author} {\bibfnamefont {J.~A.}\ \bibnamefont {Rodriguez-Rivera}}, \bibinfo {author} {\bibfnamefont {T.}~\bibnamefont {Guidi}}, \bibinfo {author} {\bibfnamefont {E.~M.}\ \bibnamefont {Wheeler}}, \bibinfo {author} {\bibfnamefont {G.~G.}\ \bibnamefont {Simeoni}}, \ and\ \bibinfo {author} {\bibfnamefont {H.}~\bibnamefont {Ryll}},\ }\bibfield  {title} {\enquote {\bibinfo {title} {Magnetic hamiltonian and phase diagram of the quantum spin liquid {Ca}$_{10}${⁢Cr}$_7${⁢O}$_{28}$},}\ }\href {https://journals.aps.org/prb/abstract/10.1103/PhysRevB.95.174414} {\bibfield  {journal} {\bibinfo  {journal} {Phys. Rev. B}\ }\textbf {\bibinfo {volume} {95}},\ \bibinfo {pages} {174414} (\bibinfo {year} {2017}{\natexlab{a}})}\BibitemShut
  {NoStop}%
\bibitem [{\citenamefont {Balz}\ \emph {et~al.}(2017{\natexlab{b}})\citenamefont {Balz}, \citenamefont {Lake}, \citenamefont {Reehuis}, \citenamefont {Islam}, \citenamefont {Singh}, \citenamefont {Prokhnenko}, \citenamefont {Pattison},\ and\ \citenamefont {Toth}}]{Balz3}%
  \BibitemOpen
  \bibfield  {author} {\bibinfo {author} {\bibfnamefont {C.}~\bibnamefont {Balz}}, \bibinfo {author} {\bibfnamefont {B.}~\bibnamefont {Lake}}, \bibinfo {author} {\bibfnamefont {M.}~\bibnamefont {Reehuis}}, \bibinfo {author} {\bibfnamefont {A.~T. M.~N.}\ \bibnamefont {Islam}}, \bibinfo {author} {\bibfnamefont {Y.}~\bibnamefont {Singh}}, \bibinfo {author} {\bibfnamefont {O.}~\bibnamefont {Prokhnenko}}, \bibinfo {author} {\bibfnamefont {P.}~\bibnamefont {Pattison}}, \ and\ \bibinfo {author} {\bibfnamefont {S.}~\bibnamefont {Toth}},\ }\bibfield  {title} {\enquote {\bibinfo {title} {Crystal growth, structure and magnetic properties of {Ca}$_{10}${⁢Cr}$_7${⁢O}$_{28}$},}\ }\href {https://iopscience.iop.org/article/10.1088/1361-648X/aa68eb} {\bibfield  {journal} {\bibinfo  {journal} {J. Phys.: Condens. Matter.}\ }\textbf {\bibinfo {volume} {29}},\ \bibinfo {pages} {225802} (\bibinfo {year} {2017}{\natexlab{b}})}\BibitemShut {NoStop}%
\bibitem [{\citenamefont {Balodhi}\ and\ \citenamefont {Singh}(2017)}]{Balodhi}%
  \BibitemOpen
  \bibfield  {author} {\bibinfo {author} {\bibfnamefont {A.}~\bibnamefont {Balodhi}}\ and\ \bibinfo {author} {\bibfnamefont {Y.}~\bibnamefont {Singh}},\ }\bibfield  {title} {\enquote {\bibinfo {title} {Synthesis and pressure and field-dependent magnetic properties of the kagome-bilayer spin liquid {Ca}$_{10}${⁢Cr}$_7${⁢O}$_{28}$},}\ }\href {https://journals.aps.org/prmaterials/abstract/10.1103/PhysRevMaterials.1.024407} {\bibfield  {journal} {\bibinfo  {journal} {Phys. Rev. Materials}\ }\textbf {\bibinfo {volume} {1}},\ \bibinfo {pages} {024407} (\bibinfo {year} {2017})}\BibitemShut {NoStop}%
\bibitem [{\citenamefont {Pohle}\ \emph {et~al.}(2017)\citenamefont {Pohle}, \citenamefont {Yang},\ and\ \citenamefont {Shannon}}]{Pohle}%
  \BibitemOpen
  \bibfield  {author} {\bibinfo {author} {\bibfnamefont {R.}~\bibnamefont {Pohle}}, \bibinfo {author} {\bibfnamefont {H.}~\bibnamefont {Yang}}, \ and\ \bibinfo {author} {\bibfnamefont {N.}~\bibnamefont {Shannon}},\ }\bibfield  {title} {\enquote {\bibinfo {title} {How many spin liquids are there in {Ca}$_{10}${Cr}$_7${O}$_{28}$?}}\ }\href {https://arxiv.org/abs/1711.03778} {\bibfield  {journal} {\bibinfo  {journal} {arXiv:1711.03778}\ } (\bibinfo {year} {2017})}\BibitemShut {NoStop}%
\bibitem [{\citenamefont {Ni}\ \emph {et~al.}(2018)\citenamefont {Ni}, \citenamefont {Liu}, \citenamefont {Yu}, \citenamefont {Cheng}, \citenamefont {Huang}, \citenamefont {Liu}, \citenamefont {Wang}, \citenamefont {Sui},\ and\ \citenamefont {Li}}]{Ni}%
  \BibitemOpen
  \bibfield  {author} {\bibinfo {author} {\bibfnamefont {J.~M.}\ \bibnamefont {Ni}}, \bibinfo {author} {\bibfnamefont {Q.~Y.}\ \bibnamefont {Liu}}, \bibinfo {author} {\bibfnamefont {Y.~J.}\ \bibnamefont {Yu}}, \bibinfo {author} {\bibfnamefont {E.~J.}\ \bibnamefont {Cheng}}, \bibinfo {author} {\bibfnamefont {Y.~Y.}\ \bibnamefont {Huang}}, \bibinfo {author} {\bibfnamefont {Z.~Y.}\ \bibnamefont {Liu}}, \bibinfo {author} {\bibfnamefont {X.~J.}\ \bibnamefont {Wang}}, \bibinfo {author} {\bibfnamefont {Y.}~\bibnamefont {Sui}}, \ and\ \bibinfo {author} {\bibfnamefont {S.~Y.}\ \bibnamefont {Li}},\ }\bibfield  {title} {\enquote {\bibinfo {title} {Ultralow-temperature heat transport in the quantum spin liquid candidate {Ca}$_{10}${⁢Cr}$_7${⁢O}$_{28}$ with a bilayer kagome lattice},}\ }\href {https://journals.aps.org/prb/abstract/10.1103/PhysRevB.97.104413} {\bibfield  {journal} {\bibinfo  {journal} {Phys. Rev. B}\ }\textbf {\bibinfo {volume} {97}},\ \bibinfo {pages} {104413} (\bibinfo {year} {2018})}\BibitemShut
  {NoStop}%
\bibitem [{\citenamefont {Sonnenschein}\ \emph {et~al.}(2019)\citenamefont {Sonnenschein}, \citenamefont {Balz}, \citenamefont {Tutsch}, \citenamefont {Lang}, \citenamefont {Ryll}, \citenamefont {Rodriguez-Rivera}, \citenamefont {Islam}, \citenamefont {Lake},\ and\ \citenamefont {Reuther}}]{Sonnenschein}%
  \BibitemOpen
  \bibfield  {author} {\bibinfo {author} {\bibfnamefont {J.}~\bibnamefont {Sonnenschein}}, \bibinfo {author} {\bibfnamefont {C.}~\bibnamefont {Balz}}, \bibinfo {author} {\bibfnamefont {U.}~\bibnamefont {Tutsch}}, \bibinfo {author} {\bibfnamefont {M.}~\bibnamefont {Lang}}, \bibinfo {author} {\bibfnamefont {H.}~\bibnamefont {Ryll}}, \bibinfo {author} {\bibfnamefont {J.~A.}\ \bibnamefont {Rodriguez-Rivera}}, \bibinfo {author} {\bibfnamefont {A.~T. M.~N.}\ \bibnamefont {Islam}}, \bibinfo {author} {\bibfnamefont {B.}~\bibnamefont {Lake}}, \ and\ \bibinfo {author} {\bibfnamefont {J.}~\bibnamefont {Reuther}},\ }\bibfield  {title} {\enquote {\bibinfo {title} {Signatures for spinons in the quantum spin liquid candidate {Ca}$_{10}${Cr}$_7${O}$_{28}$},}\ }\href {https://journals.aps.org/prb/abstract/10.1103/PhysRevB.100.174428} {\bibfield  {journal} {\bibinfo  {journal} {Phys. Rev. B}\ }\textbf {\bibinfo {volume} {100}},\ \bibinfo {pages} {174428} (\bibinfo {year} {2019})}\BibitemShut {NoStop}%
\bibitem [{\citenamefont {Kshetrimayum}\ \emph {et~al.}(2020)\citenamefont {Kshetrimayum}, \citenamefont {Balz}, \citenamefont {Lake},\ and\ \citenamefont {Eisert}}]{Kshetrimayum}%
  \BibitemOpen
  \bibfield  {author} {\bibinfo {author} {\bibfnamefont {A.}~\bibnamefont {Kshetrimayum}}, \bibinfo {author} {\bibfnamefont {C.}~\bibnamefont {Balz}}, \bibinfo {author} {\bibfnamefont {B.}~\bibnamefont {Lake}}, \ and\ \bibinfo {author} {\bibfnamefont {J.}~\bibnamefont {Eisert}},\ }\bibfield  {title} {\enquote {\bibinfo {title} {Tensor network investigation of the double layer kagome compound {Ca}$_{10}${Cr}$_7${O}$_{28}$},}\ }\href {https://www.sciencedirect.com/science/article/abs/pii/S0003491620302268} {\bibfield  {journal} {\bibinfo  {journal} {Ann. Phys.}\ }\textbf {\bibinfo {volume} {421}},\ \bibinfo {pages} {168292} (\bibinfo {year} {2020})}\BibitemShut {NoStop}%
\bibitem [{not()}]{note}%
  \BibitemOpen
  \href@noop {} {\ }\bibinfo {note} {The index for all exchange interactions is from Ref. 21.}\BibitemShut {Stop}%
\bibitem [{\citenamefont {Takahashi}\ \emph {et~al.}(2024)\citenamefont {Takahashi}, \citenamefont {Hsu}, \citenamefont {Jerzembeck}, \citenamefont {Murphy}, \citenamefont {Ward}, \citenamefont {Enright}, \citenamefont {Knapp}, \citenamefont {Puphal}, \citenamefont {Isobe}, \citenamefont {Matsumoto}, \citenamefont {Takagi},\ and\ \citenamefont {J.~C. Séamus Davis~and}}]{Blundell}%
  \BibitemOpen
  \bibfield  {author} {\bibinfo {author} {\bibfnamefont {H.}~\bibnamefont {Takahashi}}, \bibinfo {author} {\bibfnamefont {C.-C.}\ \bibnamefont {Hsu}}, \bibinfo {author} {\bibfnamefont {F.}~\bibnamefont {Jerzembeck}}, \bibinfo {author} {\bibfnamefont {J.}~\bibnamefont {Murphy}}, \bibinfo {author} {\bibfnamefont {J.}~\bibnamefont {Ward}}, \bibinfo {author} {\bibfnamefont {J.~D.}\ \bibnamefont {Enright}}, \bibinfo {author} {\bibfnamefont {J.}~\bibnamefont {Knapp}}, \bibinfo {author} {\bibfnamefont {P.}~\bibnamefont {Puphal}}, \bibinfo {author} {\bibfnamefont {M.}~\bibnamefont {Isobe}}, \bibinfo {author} {\bibfnamefont {Y.}~\bibnamefont {Matsumoto}}, \bibinfo {author} {\bibfnamefont {H.}~\bibnamefont {Takagi}}, \ and\ \bibinfo {author} {\bibfnamefont {S.~J.~Blundell}\ \bibnamefont {J.~C. Séamus Davis~and}},\ }\bibfield  {title} {\enquote {\bibinfo {title} {Spiral spin liquid noise},}\ }\href {https://arxiv.org/abs/2405.02075} {\bibfield  {journal} {\bibinfo  {journal} {arXiv:2405.02075}\ } (\bibinfo {year}
  {2024})}\BibitemShut {NoStop}%
\bibitem [{\citenamefont {Comba}\ \emph {et~al.}(2015)\citenamefont {Comba}, \citenamefont {Gro{\ss}hauser}, \citenamefont {Klingeler}, \citenamefont {Koo}, \citenamefont {Lan}, \citenamefont {M{\"u}ller}, \citenamefont {Park}, \citenamefont {Powell}, \citenamefont {Riley},\ and\ \citenamefont {Wadepohl}}]{Comba2015}%
  \BibitemOpen
  \bibfield  {author} {\bibinfo {author} {\bibfnamefont {P.}~\bibnamefont {Comba}}, \bibinfo {author} {\bibfnamefont {M.}~\bibnamefont {Gro{\ss}hauser}}, \bibinfo {author} {\bibfnamefont {R.}~\bibnamefont {Klingeler}}, \bibinfo {author} {\bibfnamefont {C.}~\bibnamefont {Koo}}, \bibinfo {author} {\bibfnamefont {Y.}~\bibnamefont {Lan}}, \bibinfo {author} {\bibfnamefont {D.}~\bibnamefont {M{\"u}ller}}, \bibinfo {author} {\bibfnamefont {J.}~\bibnamefont {Park}}, \bibinfo {author} {\bibfnamefont {A.}~\bibnamefont {Powell}}, \bibinfo {author} {\bibfnamefont {M.~J.}\ \bibnamefont {Riley}}, \ and\ \bibinfo {author} {\bibfnamefont {H.}~\bibnamefont {Wadepohl}},\ }\bibfield  {title} {\enquote {\bibinfo {title} {Magnetic interactions in a series of homodinuclear lanthanide complexes},}\ }\href {https://pubs.acs.org/doi/10.1021/acs.inorgchem.5b01673} {\bibfield  {journal} {\bibinfo  {journal} {Inorg. Chem.}\ }\textbf {\bibinfo {volume} {54}},\ \bibinfo {pages} {11247} (\bibinfo {year} {2015})}\BibitemShut {NoStop}%
\bibitem [{\citenamefont {Wellm}\ \emph {et~al.}(2020)\citenamefont {Wellm}, \citenamefont {Zeisner}, \citenamefont {Alfonsov}, \citenamefont {Sturza}, \citenamefont {Bastien}, \citenamefont {Ga{\ss}}, \citenamefont {Wurmehl}, \citenamefont {Wolter}, \citenamefont {B{\"u}chner},\ and\ \citenamefont {Kataev}}]{Wellm}%
  \BibitemOpen
  \bibfield  {author} {\bibinfo {author} {\bibfnamefont {C.}~\bibnamefont {Wellm}}, \bibinfo {author} {\bibfnamefont {J.}~\bibnamefont {Zeisner}}, \bibinfo {author} {\bibfnamefont {A.}~\bibnamefont {Alfonsov}}, \bibinfo {author} {\bibfnamefont {M.-I.}\ \bibnamefont {Sturza}}, \bibinfo {author} {\bibfnamefont {G.}~\bibnamefont {Bastien}}, \bibinfo {author} {\bibfnamefont {S.}~\bibnamefont {Ga{\ss}}}, \bibinfo {author} {\bibfnamefont {S.}~\bibnamefont {Wurmehl}}, \bibinfo {author} {\bibfnamefont {A.~U.~B.}\ \bibnamefont {Wolter}}, \bibinfo {author} {\bibfnamefont {B.}~\bibnamefont {B{\"u}chner}}, \ and\ \bibinfo {author} {\bibfnamefont {V.}~\bibnamefont {Kataev}},\ }\bibfield  {title} {\enquote {\bibinfo {title} {Magnetic interactions in the tripod kagome antiferromagnet {Mg}$_2${Gd}$_3${Sb}$_3${O}$_{14}$ probed by static magnetometry and high-field {ESR} spectroscopy},}\ }\href {https://journals.aps.org/prb/abstract/10.1103/PhysRevB.102.214414} {\bibfield  {journal} {\bibinfo  {journal} {Phys. Rev. B}\
  }\textbf {\bibinfo {volume} {102}},\ \bibinfo {pages} {214414} (\bibinfo {year} {2020})}\BibitemShut {NoStop}%
\bibitem [{\citenamefont {Cage}\ \emph {et~al.}(1997)\citenamefont {Cage}, \citenamefont {Hassan}, \citenamefont {Pardi}, \citenamefont {Krzystek}, \citenamefont {Brunel},\ and\ \citenamefont {Dalal}}]{Cage}%
  \BibitemOpen
  \bibfield  {author} {\bibinfo {author} {\bibfnamefont {B.}~\bibnamefont {Cage}}, \bibinfo {author} {\bibfnamefont {A.~K.}\ \bibnamefont {Hassan}}, \bibinfo {author} {\bibfnamefont {L.}~\bibnamefont {Pardi}}, \bibinfo {author} {\bibfnamefont {J.}~\bibnamefont {Krzystek}}, \bibinfo {author} {\bibfnamefont {L.~C.}\ \bibnamefont {Brunel}}, \ and\ \bibinfo {author} {\bibfnamefont {N.~S.}\ \bibnamefont {Dalal}},\ }\bibfield  {title} {\enquote {\bibinfo {title} {375 {GHz} {EPR} measurements on undiluted {Cr(V)} salts. the role of exchange effects and {g}-strain broadening in determining resolution in high-field {EPR} spectroscopy of {S}=1/2 paramagnets},}\ }\href {https://www.sciencedirect.com/science/article/abs/pii/S1090780796910654} {\bibfield  {journal} {\bibinfo  {journal} {J. Magn. Reson.}\ }\textbf {\bibinfo {volume} {124}},\ \bibinfo {pages} {495} (\bibinfo {year} {1997})}\BibitemShut {NoStop}%
\bibitem [{\citenamefont {Nagata}\ and\ \citenamefont {Tazuke}(1972)}]{Nagata1972}%
  \BibitemOpen
  \bibfield  {author} {\bibinfo {author} {\bibfnamefont {K.}~\bibnamefont {Nagata}}\ and\ \bibinfo {author} {\bibfnamefont {Y.}~\bibnamefont {Tazuke}},\ }\bibfield  {title} {\enquote {\bibinfo {title} {Short range order effects on {EPR} frequencies in heisenberg linear chain antiferromagnets},}\ }\href {https://journals.jps.jp/doi/10.1143/JPSJ.32.337} {\bibfield  {journal} {\bibinfo  {journal} {J. Phys. Soc. Jpn.}\ }\textbf {\bibinfo {volume} {32}},\ \bibinfo {pages} {337} (\bibinfo {year} {1972})}\BibitemShut {NoStop}%
\bibitem [{\citenamefont {Yamada}\ \emph {et~al.}(1983)\citenamefont {Yamada}, \citenamefont {Nagano},\ and\ \citenamefont {Shimoda}}]{Yamada}%
  \BibitemOpen
  \bibfield  {author} {\bibinfo {author} {\bibfnamefont {I.}~\bibnamefont {Yamada}}, \bibinfo {author} {\bibfnamefont {S.}~\bibnamefont {Nagano}}, \ and\ \bibinfo {author} {\bibfnamefont {S.}~\bibnamefont {Shimoda}},\ }\bibfield  {title} {\enquote {\bibinfo {title} {Temperature dependences of {EPR} fields in {K}$_2${Cu}$_x${Mn}$_{1-x}${F}$_4$},}\ }\href {https://www.sciencedirect.com/science/article/pii/0378436383900098} {\bibfield  {journal} {\bibinfo  {journal} {Phyica B+C}\ }\textbf {\bibinfo {volume} {123}},\ \bibinfo {pages} {47} (\bibinfo {year} {1983})}\BibitemShut {NoStop}%
\bibitem [{\citenamefont {Kubo}\ and\ \citenamefont {Tomit}(1954)}]{Kubo}%
  \BibitemOpen
  \bibfield  {author} {\bibinfo {author} {\bibfnamefont {R.}~\bibnamefont {Kubo}}\ and\ \bibinfo {author} {\bibfnamefont {K.}~\bibnamefont {Tomit}},\ }\bibfield  {title} {\enquote {\bibinfo {title} {A general theory of magnetic resonance absorption},}\ }\href {https://journals.jps.jp/doi/abs/10.1143/JPSJ.9.888} {\bibfield  {journal} {\bibinfo  {journal} {J. Phys. Soc. Jpn.}\ }\textbf {\bibinfo {volume} {9}},\ \bibinfo {pages} {888} (\bibinfo {year} {1954})}\BibitemShut {NoStop}%
\bibitem [{\citenamefont {Oshikawa}\ and\ \citenamefont {Affleck}(2002)}]{Oshikawa}%
  \BibitemOpen
  \bibfield  {author} {\bibinfo {author} {\bibfnamefont {M.}~\bibnamefont {Oshikawa}}\ and\ \bibinfo {author} {\bibfnamefont {I.}~\bibnamefont {Affleck}},\ }\bibfield  {title} {\enquote {\bibinfo {title} {Electron spin resonance in {S} = 1/2 antiferromagnetic chains},}\ }\href {https://journals.aps.org/prb/abstract/10.1103/PhysRevB.65.134410} {\bibfield  {journal} {\bibinfo  {journal} {Phys. Rev. B}\ }\textbf {\bibinfo {volume} {65}},\ \bibinfo {pages} {134410} (\bibinfo {year} {2002})}\BibitemShut {NoStop}%
\bibitem [{\citenamefont {Choukroun}\ \emph {et~al.}(2001)\citenamefont {Choukroun}, \citenamefont {Richard},\ and\ \citenamefont {Stepanov}}]{Choukroun}%
  \BibitemOpen
  \bibfield  {author} {\bibinfo {author} {\bibfnamefont {J.}~\bibnamefont {Choukroun}}, \bibinfo {author} {\bibfnamefont {J.-L.}\ \bibnamefont {Richard}}, \ and\ \bibinfo {author} {\bibfnamefont {A.}~\bibnamefont {Stepanov}},\ }\bibfield  {title} {\enquote {\bibinfo {title} {High-temperature electron paramagnetic resonance in magnets with the {Dzyaloshinskii-Moriya} interaction},}\ }\href {https://journals.aps.org/prl/abstract/10.1103/PhysRevLett.87.127207} {\bibfield  {journal} {\bibinfo  {journal} {Phys. Rev. Lett.}\ }\textbf {\bibinfo {volume} {87}},\ \bibinfo {pages} {127207} (\bibinfo {year} {2001})}\BibitemShut {NoStop}%
\bibitem [{\citenamefont {Dalal}\ \emph {et~al.}(1981)\citenamefont {Dalal}, \citenamefont {Millar}, \citenamefont {Jagadeesh},\ and\ \citenamefont {Seehra}}]{Dalal1981}%
  \BibitemOpen
  \bibfield  {author} {\bibinfo {author} {\bibfnamefont {N.~S.}\ \bibnamefont {Dalal}}, \bibinfo {author} {\bibfnamefont {J.~M.}\ \bibnamefont {Millar}}, \bibinfo {author} {\bibfnamefont {M.~S.}\ \bibnamefont {Jagadeesh}}, \ and\ \bibinfo {author} {\bibfnamefont {M.~S.}\ \bibnamefont {Seehra}},\ }\bibfield  {title} {\enquote {\bibinfo {title} {Paramagnetic resonance, magnetic susceptibility, and antiferromagnetic exchange in a {Cr}$^{5+}$ paramagnet: Potassium perchromate {(K}$_3${CrO}$_8${)}},}\ }\href {https://pubs.aip.org/aip/jcp/article-abstract/74/3/1916/90372/Paramagnetic-resonance-magnetic-susceptibility-and?redirectedFrom=fulltext} {\bibfield  {journal} {\bibinfo  {journal} {J. Chem. Phys.}\ }\textbf {\bibinfo {volume} {74}},\ \bibinfo {pages} {1916} (\bibinfo {year} {1981})}\BibitemShut {NoStop}%
\bibitem [{\citenamefont {Huber}\ and\ \citenamefont {Seehra}(1975)}]{Huber}%
  \BibitemOpen
  \bibfield  {author} {\bibinfo {author} {\bibfnamefont {D.}~\bibnamefont {Huber}}\ and\ \bibinfo {author} {\bibfnamefont {M.}~\bibnamefont {Seehra}},\ }\bibfield  {title} {\enquote {\bibinfo {title} {Contribution of the spin-phonon interaction to the paramagnetic resonance linewidth of {CrBr}$_3$},}\ }\href {https://www.sciencedirect.com/science/article/pii/0022369775900943} {\bibfield  {journal} {\bibinfo  {journal} {J. Phys. Chem. Solids}\ }\textbf {\bibinfo {volume} {36}},\ \bibinfo {pages} {723} (\bibinfo {year} {1975})}\BibitemShut {NoStop}%
\bibitem [{\citenamefont {Heinrich}\ \emph {et~al.}(2003)\citenamefont {Heinrich}, \citenamefont {von Nidda}, \citenamefont {Krimmel}, \citenamefont {Loidl}, \citenamefont {Eremina}, \citenamefont {Ineev}, \citenamefont {Kochelaev}, \citenamefont {Prokofiev},\ and\ \citenamefont {Assmus}}]{Heinrich}%
  \BibitemOpen
  \bibfield  {author} {\bibinfo {author} {\bibfnamefont {M.}~\bibnamefont {Heinrich}}, \bibinfo {author} {\bibfnamefont {H.-A.~Krug}\ \bibnamefont {von Nidda}}, \bibinfo {author} {\bibfnamefont {A.}~\bibnamefont {Krimmel}}, \bibinfo {author} {\bibfnamefont {A.}~\bibnamefont {Loidl}}, \bibinfo {author} {\bibfnamefont {R.M.}\ \bibnamefont {Eremina}}, \bibinfo {author} {\bibfnamefont {A.~D.}\ \bibnamefont {Ineev}}, \bibinfo {author} {\bibfnamefont {B.~I.}\ \bibnamefont {Kochelaev}}, \bibinfo {author} {\bibfnamefont {A.~V.}\ \bibnamefont {Prokofiev}}, \ and\ \bibinfo {author} {\bibfnamefont {W.}~\bibnamefont {Assmus}},\ }\bibfield  {title} {\enquote {\bibinfo {title} {Structural and magnetic properties of {CuSb}$_2${⁢O}$_6$ probed by {ESR}},}\ }\href {https://journals.aps.org/prb/abstract/10.1103/PhysRevB.67.224418} {\bibfield  {journal} {\bibinfo  {journal} {Phys. Rev. B}\ }\textbf {\bibinfo {volume} {67}},\ \bibinfo {pages} {224418} (\bibinfo {year} {2003})}\BibitemShut {NoStop}%
\bibitem [{\citenamefont {Eremin}\ \emph {et~al.}(2008)\citenamefont {Eremin}, \citenamefont {Zakharov}, \citenamefont {von Nidda}, \citenamefont {Eremina}, \citenamefont {Shuvaev}, \citenamefont {Pimenov}, \citenamefont {Ghigna}, \citenamefont {Deisenhofer},\ and\ \citenamefont {Loidl}}]{Eremin2008}%
  \BibitemOpen
  \bibfield  {author} {\bibinfo {author} {\bibfnamefont {M.V.}\ \bibnamefont {Eremin}}, \bibinfo {author} {\bibfnamefont {D.V.}\ \bibnamefont {Zakharov}}, \bibinfo {author} {\bibfnamefont {H.-A.~Krug}\ \bibnamefont {von Nidda}}, \bibinfo {author} {\bibfnamefont {R.M.}\ \bibnamefont {Eremina}}, \bibinfo {author} {\bibfnamefont {A.}~\bibnamefont {Shuvaev}}, \bibinfo {author} {\bibfnamefont {A.}~\bibnamefont {Pimenov}}, \bibinfo {author} {\bibfnamefont {P.}~\bibnamefont {Ghigna}}, \bibinfo {author} {\bibfnamefont {J.}~\bibnamefont {Deisenhofer}}, \ and\ \bibinfo {author} {\bibfnamefont {A.}~\bibnamefont {Loidl}},\ }\bibfield  {title} {\enquote {\bibinfo {title} {Dynamical dzyaloshinsky-moriya interaction in {KCuF}$_3$},}\ }\href {https://journals.aps.org/prl/abstract/10.1103/PhysRevLett.101.147601} {\bibfield  {journal} {\bibinfo  {journal} {Phys. Rev. Lett.}\ }\textbf {\bibinfo {volume} {101}},\ \bibinfo {pages} {147601} (\bibinfo {year} {2008})}\BibitemShut {NoStop}%
\bibitem [{\citenamefont {Werner}\ \emph {et~al.}(2016)\citenamefont {Werner}, \citenamefont {Koo}, \citenamefont {Klingeler}, \citenamefont {Vasiliev}, \citenamefont {Ovchenkov}, \citenamefont {Polovkova}, \citenamefont {Raganyan},\ and\ \citenamefont {Zvereva}}]{Werner2016}%
  \BibitemOpen
  \bibfield  {author} {\bibinfo {author} {\bibfnamefont {J.}~\bibnamefont {Werner}}, \bibinfo {author} {\bibfnamefont {C.}~\bibnamefont {Koo}}, \bibinfo {author} {\bibfnamefont {R.}~\bibnamefont {Klingeler}}, \bibinfo {author} {\bibfnamefont {A.~N.}\ \bibnamefont {Vasiliev}}, \bibinfo {author} {\bibfnamefont {Y.~A.}\ \bibnamefont {Ovchenkov}}, \bibinfo {author} {\bibfnamefont {A.~S.}\ \bibnamefont {Polovkova}}, \bibinfo {author} {\bibfnamefont {G.~V.}\ \bibnamefont {Raganyan}}, \ and\ \bibinfo {author} {\bibfnamefont {E.~A.}\ \bibnamefont {Zvereva}},\ }\bibfield  {title} {\enquote {\bibinfo {title} {Magnetic anisotropy and the phase diagram of chiral {MnSb}$_2${⁢O}$_6$},}\ }\href {https://journals.aps.org/prb/abstract/10.1103/PhysRevB.94.104408} {\bibfield  {journal} {\bibinfo  {journal} {Phys. Rev. B}\ }\textbf {\bibinfo {volume} {94}},\ \bibinfo {pages} {104408} (\bibinfo {year} {2016})}\BibitemShut {NoStop}%
\bibitem [{\citenamefont {Zorko}\ \emph {et~al.}(2008)\citenamefont {Zorko}, \citenamefont {Nellutla}, \citenamefont {van Tol}, \citenamefont {Brunel}, \citenamefont {Bert}, \citenamefont {Duc}, \citenamefont {Trombe}, \citenamefont {de~Vries}, \citenamefont {Harrison},\ and\ \citenamefont {Mendels}}]{Zorko}%
  \BibitemOpen
  \bibfield  {author} {\bibinfo {author} {\bibfnamefont {A.}~\bibnamefont {Zorko}}, \bibinfo {author} {\bibfnamefont {S.}~\bibnamefont {Nellutla}}, \bibinfo {author} {\bibfnamefont {J.}~\bibnamefont {van Tol}}, \bibinfo {author} {\bibfnamefont {L.~C.}\ \bibnamefont {Brunel}}, \bibinfo {author} {\bibfnamefont {F.}~\bibnamefont {Bert}}, \bibinfo {author} {\bibfnamefont {F.}~\bibnamefont {Duc}}, \bibinfo {author} {\bibfnamefont {J.-C.}\ \bibnamefont {Trombe}}, \bibinfo {author} {\bibfnamefont {M.~A.}\ \bibnamefont {de~Vries}}, \bibinfo {author} {\bibfnamefont {A.}~\bibnamefont {Harrison}}, \ and\ \bibinfo {author} {\bibfnamefont {P.}~\bibnamefont {Mendels}},\ }\bibfield  {title} {\enquote {\bibinfo {title} {Dzyaloshinsky-moriya anisotropy in the spin-1/2 kagome compound {ZnCu}$_3${⁢(OH)}$_6${⁢Cl}$_2$},}\ }\href {https://journals.aps.org/prl/abstract/10.1103/PhysRevLett.101.026405} {\bibfield  {journal} {\bibinfo  {journal} {Phys. Rev. Lett.}\ }\textbf {\bibinfo {volume} {101}},\ \bibinfo {pages} {026405}
  (\bibinfo {year} {2008})}\BibitemShut {NoStop}%
\bibitem [{\citenamefont {Zhang}\ \emph {et~al.}(2010)\citenamefont {Zhang}, \citenamefont {Ohta}, \citenamefont {Okubo}, \citenamefont {Fujisawa}, \citenamefont {Sakurai}, \citenamefont {Okamoto}, \citenamefont {Yoshida},\ and\ \citenamefont {Hiroi}}]{Zhang2010}%
  \BibitemOpen
  \bibfield  {author} {\bibinfo {author} {\bibfnamefont {W.}~\bibnamefont {Zhang}}, \bibinfo {author} {\bibfnamefont {H.}~\bibnamefont {Ohta}}, \bibinfo {author} {\bibfnamefont {S.}~\bibnamefont {Okubo}}, \bibinfo {author} {\bibfnamefont {M.}~\bibnamefont {Fujisawa}}, \bibinfo {author} {\bibfnamefont {T.}~\bibnamefont {Sakurai}}, \bibinfo {author} {\bibfnamefont {Y.}~\bibnamefont {Okamoto}}, \bibinfo {author} {\bibfnamefont {H.}~\bibnamefont {Yoshida}}, \ and\ \bibinfo {author} {\bibfnamefont {Z.}~\bibnamefont {Hiroi}},\ }\bibfield  {title} {\enquote {\bibinfo {title} {High-field {ESR} measurements of {S}=1/2 kagome lattice antiferromagnet {BaCu}$_3${V}$_2${O}$_8${(OH)}$_2$},}\ }\href {https://journals.jps.jp/doi/10.1143/JPSJ.79.023708} {\bibfield  {journal} {\bibinfo  {journal} {J. Phys. Soc. Jpn.}\ }\textbf {\bibinfo {volume} {79}},\ \bibinfo {pages} {023708} (\bibinfo {year} {2010})}\BibitemShut {NoStop}%
\bibitem [{\citenamefont {Zorko}\ \emph {et~al.}(2013)\citenamefont {Zorko}, \citenamefont {Bert}, \citenamefont {Ozarowski}, \citenamefont {van Tol}, \citenamefont {Boldrin}, \citenamefont {Wills},\ and\ \citenamefont {Mendels}}]{Zorko2013}%
  \BibitemOpen
  \bibfield  {author} {\bibinfo {author} {\bibfnamefont {A.}~\bibnamefont {Zorko}}, \bibinfo {author} {\bibfnamefont {F.}~\bibnamefont {Bert}}, \bibinfo {author} {\bibfnamefont {A.}~\bibnamefont {Ozarowski}}, \bibinfo {author} {\bibfnamefont {J.}~\bibnamefont {van Tol}}, \bibinfo {author} {\bibfnamefont {D.}~\bibnamefont {Boldrin}}, \bibinfo {author} {\bibfnamefont {A.~S.}\ \bibnamefont {Wills}}, \ and\ \bibinfo {author} {\bibfnamefont {P.}~\bibnamefont {Mendels}},\ }\bibfield  {title} {\enquote {\bibinfo {title} {Dzyaloshinsky-moriya interaction in vesignieite: A route to freezing in a quantum kagome antiferromagnet},}\ }\href {https://journals.aps.org/prb/abstract/10.1103/PhysRevB.88.144419} {\bibfield  {journal} {\bibinfo  {journal} {Phys. Rev. B}\ }\textbf {\bibinfo {volume} {88}},\ \bibinfo {pages} {144419} (\bibinfo {year} {2013})}\BibitemShut {NoStop}%
\bibitem [{\citenamefont {Choi}\ \emph {et~al.}(2012)\citenamefont {Choi}, \citenamefont {Wang}, \citenamefont {Ozarowski}, \citenamefont {van Tol}, \citenamefont {Zhou}, \citenamefont {Wiebe}, \citenamefont {Skoursk},\ and\ \citenamefont {Dalal}}]{Choi2012}%
  \BibitemOpen
  \bibfield  {author} {\bibinfo {author} {\bibfnamefont {K.~Y.}\ \bibnamefont {Choi}}, \bibinfo {author} {\bibfnamefont {Z.}~\bibnamefont {Wang}}, \bibinfo {author} {\bibfnamefont {A.}~\bibnamefont {Ozarowski}}, \bibinfo {author} {\bibfnamefont {J.}~\bibnamefont {van Tol}}, \bibinfo {author} {\bibfnamefont {H.~D.}\ \bibnamefont {Zhou}}, \bibinfo {author} {\bibfnamefont {C.~R.}\ \bibnamefont {Wiebe}}, \bibinfo {author} {\bibfnamefont {Y.}~\bibnamefont {Skoursk}}, \ and\ \bibinfo {author} {\bibfnamefont {N.~S.}\ \bibnamefont {Dalal}},\ }\bibfield  {title} {\enquote {\bibinfo {title} {Spin dynamics of the {S} = 5/2 2d triangular antiferromagnet {Ba}$_3${NbFe}$_3${Si}$_2${O}$_{14}$},}\ }\href {https://iopscience.iop.org/article/10.1088/0953-8984/24/24/246001} {\bibfield  {journal} {\bibinfo  {journal} {J. Phys.: Condens. Matter}\ }\textbf {\bibinfo {volume} {24}},\ \bibinfo {pages} {246001} (\bibinfo {year} {2012})}\BibitemShut {NoStop}%
\end{thebibliography}%

\end{document}